\documentclass[12pt]{article}
\pdfoutput=1
\usepackage[T1]{fontenc}
\usepackage{amsmath,amssymb,mathrsfs}
\usepackage{paralist}
\usepackage{slashed}
\usepackage{graphicx}
\usepackage{subfigure}
\usepackage{color}
\usepackage{fullpage}
\usepackage{hyperref}
\usepackage{amscd}
\usepackage{xcolor}
\usepackage{braket}
\usepackage{amsfonts}
\usepackage{tikz}
\usepackage{cite}
\usepackage{booktabs}
\usepackage{pdfpages}
\usepackage{empheq}
\usepackage{xparse}

\catcode`\@=11 
\def\numberbysection{\@addtoreset{equation}{section} 
        \def\theequation{\thesection.\arabic{equation}}} 

\newcommand{\bs}{\boldsymbol}
\def\be{\begin{equation}} 
\def\ee{\end{equation}} 
\def\ba{\begin{eqnarray}} 
\def\ea{\end{eqnarray}} 
\def\bali{\begin{align}}
\def\eali{\end{align}}
 
\def\ov{\overline} 
 
\def\R{{\rm Re}} 
\def\Z{\mathbb{Z}}

\def\nl{\nonumber \\} 
 
\def\ra{\rangle} 
\def\la{\langle} 
\def\de{\partial} 
\def\wt{\widetilde} 
\def\wh{\widehat}

\def\a{\alpha} 
\def\b{\beta} 
\def\g{\gamma} 
\def\G{\Gamma} 
\def\D{\Delta} 
\def\d{\delta} 
 
\def\eps{\varepsilon} 
\def\z{\zeta}

\def\l{\lambda} 
\def\L{\Lambda} 
\def\m{\mu} 
\def\n{\nu}

\def\p{\pi} 
 
\def\r{\rho} 
\def\s{\sigma} 
 
\def\t{\tau} 
\def\f{\phi} 
\def\vf{\varphi} 
\def\F{\Phi} 
 
\def\w{\omega}

\def\th{\theta}

\begin{document} 
 
\begin{titlepage} 
\begin{center} 
\vskip .6 in 
{\LARGE Quantization of a Self-dual Conformal Theory}\\
\medskip
{\LARGE in $(2+1)$ Dimensions} 
\vskip 0.2in 
Francesco ANDREUCCI${}^{(a,b)}$, Andrea CAPPELLI${}^{(c)}$ and 
Lorenzo MAFFI${}^{(a,c)}$
 \medskip

{\em ${}^{(a)}$Dipartimento di Fisica, Universit\`a di Firenze\\ 
Via G. Sansone 1, 50019 Sesto Fiorentino - Firenze, Italy} \\
{\em ${}^{(b)}$SISSA, Via Bonomea 265, 34136 Trieste, Italy}\\
{\em ${}^{(c)}$INFN, Sezione di Firenze\\
Via G. Sansone 1, 50019 Sesto Fiorentino - Firenze, Italy}
\end{center} 
\vskip .2 in 

\begin{abstract} 
Compact nonlocal Abelian gauge theory in $(2+1)$ dimensions, also
known as loop model, is a massless theory with a critical line
that is explicitly covariant under duality transformations.
It corresponds to the large $N_F$ limit of self-dual
electrodynamics in mixed three-four dimensions.
It also provides a bosonic description for surface excitations of
three-dimensional topological insulators.  Upon mapping the model to a
local gauge theory in $(3+1)$ dimensions, we compute the spectrum of
electric and magnetic solitonic excitations and the partition function
on the three torus $\mathbb{T}_3$.
Analogous results for the $S^2\times S^1$ geometry show
that the theory is conformal invariant and determine
the manifestly self-dual spectrum of conformal fields,
corresponding to order-disorder excitations with fractional statistics.
\end{abstract} 
 
\vfill 
\end{titlepage} 
\pagenumbering{arabic} 
\numberbysection

 
\section{Introduction} 

The nonlocal Abelian gauge theory is defined by the following
action \cite{fradkin-loop}:
\begin{equation}
  \label{S_loop_x}
  S[a_\mu]= \frac{g}{16\pi^3}
  \int d^{3}x\, d^{3}y \, F_{\mu \nu}(x) \dfrac{1}{(x-y)^{2}} F_{\mu \nu}(y)
  + i\frac{f}{4\pi} \int d^{3}x\, \eps^{\mu\nu\r}a_\mu \de_\nu a_\r .
\end{equation}
In this expression, $F_{\mu\nu} =\de_\mu a_\nu -\de_\nu a_\mu$ and the
gauge field is assumed to be compact, $a_\mu\sim a_\mu +2\pi r n_\mu$,
with $r$ the compactification radius and $n_\mu\in \Z$.
The theory is quadratic but nontrivial
owing to its solitonic spectrum of electric and magnetic excitations.
In this work, we shall resolve the difficulties due to nonlocality
of the $1/x^2$ kernel and obtain such spectrum. There
are two coupling constants, $g$ and $f$, but most of the 
results will concern the $f=0$ case.

The action (\ref{S_loop_x}) can be rewritten in terms of 
degrees of freedom that are conserved currents,
\be
j^\mu=\frac{1}{2\pi}\eps^{\mu\nu\r}\de_\nu a_\r \ .
\label{j_curr}
\ee

Once formulated on a Euclidean lattice, it defines a statistical
model where the currents describe random loops that interact by
the potential $\int j_\mu (1/x^2) j_\mu$, giving rise to an
interesting phase diagram: in this formulation, the theory is called
`loop model'. In the following we shall mostly use this short-hand
name.

The theory has appeared in a number of recent research topics:
\begin{itemize}
\item
In the study of massless excitations at the surface of
three-dimensional topological insulators \cite{fradkin-book}
\cite{bernevig-book}. While the free fermion theory 
is well understood, the bosonic description, following
from the bulk topological gauge theory \cite{moore-BF}, is not yet
fully developed. In an earlier work \cite{CRS}, the bosonic nonlocal
action was argued to be relevant because it reproduces the fermion
dynamics in the semiclassical, low-energy limit. Upon varying the
coupling constant, this bosonic theory can also describe massless
excitations with fractional statistics, that exist at the surface of
interacting  topological insulators
\cite{fradkin-BF}.

\item 
  The boson-fermion correspondence, i.e. bosonization in $(2+1)$
  dimensions, is part of the web of duality relations that have been
  extensively analyzed in the recent years \cite{dual}
  \cite{dual-rev}. The loop model provides a neat example of a
  massless theory that is covariant under duality transformations,
  corresponding to $SL(2,\Z)$ maps of the complex coupling
  $\t=f+ig$.  In particular, the loop model is equal to
  self-dual electrodynamics in mixed dimensions ($QED_{4,3}$)
  \cite{son-F}, in the limit of large number of fermion fields
  $N_F\to\infty$.

\item
  Finally, the loop model provides a nontrivial example of a conformal
  field theory in $(2+1)$ dimensions possessing a critical line
  parameterized by the coupling constant $g$; its solitonic
  excitations correspond to order-disorder fields, generalization of
  vertex operators, with fermionic or anyonic statistical phases
  depending on the value of $g$.  These features remind of the
  compactified boson conformal theory in $(1+1)$ dimensions
  \cite{cft}, corresponding to the massless phase of the $XY$
  statistical spin model \cite{kogut}.  In our analysis, we shall
  point out similarities and differences between the two theories.
  \end{itemize}

In section two, some features of the loop models are briefly recalled
and rederived. Starting from the qualitative determination of the
phase diagram using energy-entropy Peierls estimates, we introduce the
physics at the surface of topological insulators and the solitonic
excitations that occur in these systems. Next, we show that the loop
model enjoys exact self-duality and matches the limit $N_F\to\infty$ 
of $QED_{4,3}$.

In section three, our quantization procedure is presented. Inspired
by the relation with $QED_{4,3}$, we reformulate the loop model as
ordinary electrodynamics in $(3+1)$ dimension, where the photons interact
by a BF action defined on a two-dimensional space slice. We then obtain the
solitonic spectrum by the usual analysis of nontrivial solutions
of the equations of motion. 

We consider the model on the toroidal geometry $\mathbb{T}_3\times I$,
where $I$ is the interval in the extra dimension: an infrared
cutoff is needed, that is actually a crucial aspect for the definition of the
theory.  We obtain the partition function for two choices of the
cutoff: a fixed scale $1/M$ and the spatial size of the 
torus.  In the first case, the loop model reduces on-shell to a local
theory analyzed earlier \cite{CRS}, thus providing a check of our
results; however, the mass $M$ breaks scale invariance.
The second choice of size-dependent cutoff is thus preferable
because it leads to a conformal invariant quantum theory.

In section four, the solitonic spectrum and the partition function
are determined for the geometry $S^2\times S^1$, where the dimensional
extension is obtained by considering $S^2$ as the equator of $S^3$. 
Such geometries are related to
flat space by a conformal transformation, where the Hamiltonian 
maps into the dilatation operator. Therefore, the solitonic energies
determine the spectrum of conformal dimensions of the fields.
The computation of the partition function in this geometry
explicitly confirms the conformal invariance of the theory.

In section five, we analyze our results and briefly describe the
$(2+1)$-dimensional order-disorder fields of the loop model.
In section six, we outline possible developments and conclude.
In Appendix A, we give some details on the Peierls argument
and in Appendix B we report the calculations for the partition function
on $S^2\times S^1$.

  
\section{Properties of the loop model}


\subsection{Notations}

We first write down some useful formulas and notations. 
The $(2+1)$-dimensional Euclidean Laplacian and its square root are
indicated as follows,
\be
\de_\mu^2 \equiv \de^2, \qquad\quad \sqrt{-\de^2}\equiv \de,
\label{delta_def}
\ee
and their Green functions in coordinate space are,
\begin{equation} 
\label{green_f}
\left(\dfrac{1}{-\de^{2}}\right)_{x,y} = 
\dfrac{1}{4\pi}\dfrac{1}{\sqrt{(x-y)^2}},\qquad
\quad \left(\dfrac{1}{\sqrt{-\de^2}}\right)_{x,y} = 
\dfrac{1}{2\pi^{2}}\dfrac{1}{(x-y)^{2}}\, .
\end{equation}
If follows that the loop model action (\ref{S_loop_x}) 
can be rewritten in term of the following kernel:
\ba
  S[a] &= &\dfrac{1}{4\pi}\int{d^{3}x\, d^{3}y\,
    a_{\mu}(x)\, D_{\mu \nu}(g,f)(x,y)\, a_{\nu}}(y)\, ,
  \nl
&&  D_{\mu \nu}(g,f) =  g \dfrac{1}{\de} ( -\delta_{\mu \nu} \de^{2} +
  \de_{\mu} \de_{\nu}) +if \varepsilon_{\mu \rho \nu} \de_{\rho}\, .
  \label{S_loop_ker}
\ea
This satisfies the following inversion relation \cite{fradkin-loop}:
\begin{align}
&\int{d^{3}x d^{3}y \, j_{\mu} D^{-1}(g,f)_{\mu \nu}j_{\nu}} =
  \int{d^{3}x d^{3}y \, \zeta_{\mu} D_{\mu \nu}(\wh{g}, \wh{f})
    \zeta_{\nu}}\, , \\ &\wh{g}= \dfrac{g}{g^{2}+f^{2}}, \qquad
  \wh{f}= \dfrac{-f}{g^{2}+f^{2}}, \qquad\quad j_{\mu}= \varepsilon_{\mu
    \nu \rho} \de_{\nu}\zeta_{\rho} \, ,
\label{g_f_tilde}
\end{align}
that is obtained for $\de_\mu j_\mu=\de_\mu \z_\mu=0$. This relation
will be used extensively. Note that the map (\ref{g_f_tilde}) is particularly
simple, $\wh{\t}=-1/\t$, in terms of the complex coupling constant
$\t=f+ig$.


\subsection{Phase diagram} 
\label{Peierls_loop}

In this section, we determine the phase diagram of the model by using
Peierls arguments \cite{kogut}. These amounts to estimates of the probability
$P\propto \exp(-\beta \Delta F) = \exp(-\b\D E +\D S)$ for creating a
``disorder'' excitation above the ``ordered'' ground state.  If the
energy cost $\D E$ of the excitation exceeds the entropy $\D S$
(logarithm of the multiplicity) in the thermodynamic limit,
then the excitation is suppressed and the ordered phase is
stable; otherwise the entropy wins and excitations proliferate,
leading to a disordered (massive) phase.

A well-known examples is given by the estimate of free energy for
one vortex in the massless phase of the $XY$ spin model in two
dimensions \cite{kogut}. In this case, both energy and entropy grow
logarithmically with the system size $L$, leading to $\b \D F \sim (\b -\b_c)
\log (L/a)$ ($a$ is the lattice size, the $UV$ cutoff).
One finds that the massless phase is stable for $\b>\b_c$, i.e.
$P\to 0$ for $L\to\infty$, while the disordered phase takes place for
$\b<\b_c$. The massless phase corresponds to the critical line
of the compactified boson conformal theory with central charge $c=1$.
Thanks to exact bosonization in $(1+1)$ dimensions, the bosonic
theory describes both free and interacting massless fermions at different
points of the critical line.

The loop model presents a similar behavior in one dimension higher,
with a massless phase corresponding to the critical line
$g>g_c$. In order to prove this fact,
let us consider the action (\ref{S_loop_x}), setting $f=0$ but adding a
local Yang-Mills term:
\begin{equation}
  \label{S_peierls}
  S[a_\mu]= \frac{g}{16\pi^3}
  \int d^{3}x\, d^{3}y \, F_{\mu \nu}(x) \dfrac{1}{(x-y)^{2}} F_{\mu \nu}(y)
  + \dfrac{t}{M} \int{d^{3}x\, F_{\mu \nu} F_{\mu \nu} }\, .
\end{equation}
In this expression, $g$ and $t$ are dimensionless couplings and $M$ is
a mass scale.  In absence of matter fields, the Yang-Mills term is
actually irrelevant in the renormalization-group sense.

The compact Abelian theory, say on a lattice, possesses isolated
monopole configurations (strictly speaking, they are instantons of the
three-dimensional Euclidean theory), that obey the quantization
condition:
\be
\int_{S^2} F =2\pi \frac{M_0}{q_0}, \quad M_0\in\Z ,
\label{dirac_q}
\ee
where $F$ is the gauge field two-form and $q_0$ is the minimal charge
in the theory, trade-off for the compactification radius.

The evaluation of the loop model action (\ref{S_peierls}) for one
monopole configuration of minimal magnetic charge ($M_0=1$) is
carried out in Appendix A, leading to the result:
\be
\b \D F=\frac{1}{2q_0^2}\left(\frac{g}{\pi}\log\left(\frac{L}{a}\right) +
\frac{t}{Ma}\right) -3 \log\left(\frac{L}{a}\right) .
\label{deltaF_mono}
\ee
We see that the nonlocal term yields a logarithmic energy,
while the local Yang-Mills action gives a
constant. The entropy is also logarithmic, counting
the number of lattice cubes which can host monopoles.
Therefore, in ordinary Yang-Mills theory ($g=0$),
the entropy always dominates and
monopoles proliferate: the system is disordered for any coupling.
We recover here Polyakov's result that Abelian lattice Yang-Mills theory
is massive and confines charges \cite{polyakov}.

The nonlocal term provides a completely different dynamics, allowing
for a stable massless phase without monopoles for
$g>g_c\sim 6\pi q_0^2$,
which corresponds to the critical line of the loop model.
The analogy with the $XY$ model in one lower dimension is apparent.

In the massless phase $g>g_c$, we can consider other excitations
corresponding to closed loops of flux lines.  Let us now estimate
whether large loops of length $R$ are allowed or suppressed as a
function of the couplings $g$ and $t$.  The multiplicity of loops can
be estimated as $5^{(R/a)}$ from the number of random walk steps.
Therefore the entropy is linear in $R$.

The associated activation energy is obtained by evaluating the action 
(\ref{S_peierls}) for the configuration of a line of minimal 
flux $\F_0$ directed along the $z$-axis with length $R$: this corresponds to
the field configuration $F_{12}=\F_0 \d(x)\d(y)$, for $0<z<R$.
The resulting free energy is, for large $R$,
\begin{equation}
  \label{deltaF_vort}
  \beta \Delta F  = \F_0^2\left( \frac{g}{16\pi^3}+ \frac{t}{Ma}\right)
  \left(\frac{R}{a}\right)  - \ln(5)\left(\frac{R}{a}\right)  ,
  \qquad\quad (g >g_{c})\, .
\end{equation}

We see that both the local and nonlocal terms contribute to the
energy of closed loops and that energy and entropy can balance.
The condition $\beta \Delta F=0$ defines the critical line
$t_c(g)=a-bg$, with $a,b$ positive constants, in the plane $(g,t)$:
this line separates the (massive) phase $t<t_c(g)$, in which large loops
proliferate, from the phase $t>t_c(g)$ in which loops are tiny.
Another interesting line is given by the condition of vanishing energy
(Euclidean action) $g+ct=0$, with $c$ positive constant, below which
the theory is not defined.

\begin{figure}
\begin{center}
\includegraphics[scale=1.2]{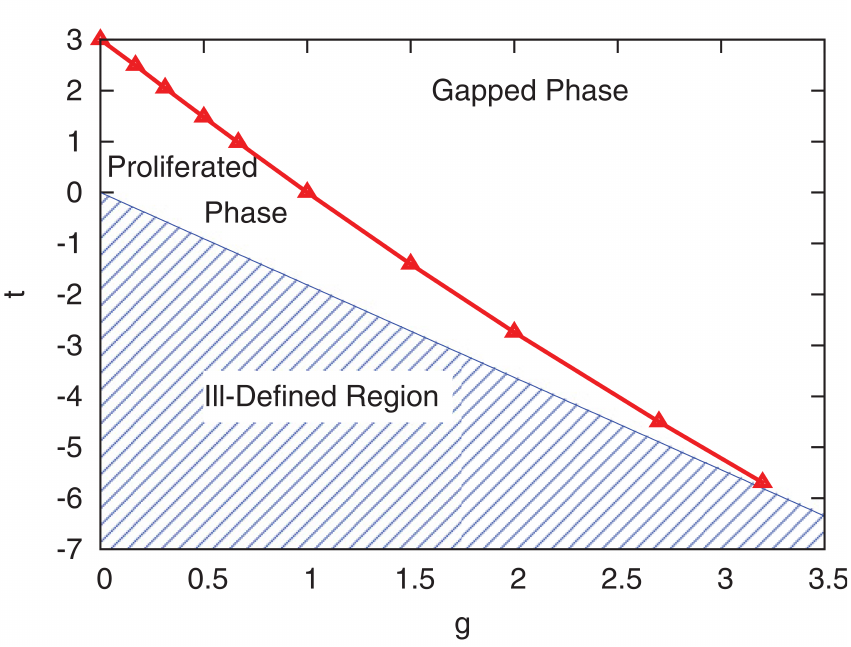}
\caption{Phase diagram of the loop model found by
numerical simulation \cite{motrunich}. The two phases
  with proliferating and small closed loops are separated
  by the critical line $t_c(g)$ drawn in red.} 
\label{loop_phase}
\end{center}
\end{figure}

The loop model with action (\ref{S_peierls}) has been simulated on a lattice in
Ref.\cite{motrunich}: Fig.\ref{loop_phase} shows the numerical
results for the phase diagram in the $(g,t)$ plane, that are in
qualitative agreement with the Peierls estimate (\ref{deltaF_vort}).
We remark that the simulation enforces the closed loop condition and
cannot see the $g<g_c$ phase of free monopoles.
We also note that the coupling $t$ is irrelevant and thus disappears
in the IR limit: therefore, in the low-energy effective action there
remains the nonlocal term and the phase diagram reduces to the 
critical line parameterized by $g>g_c$.

\subsection{Surface excitations of three-dimensional topological insulators}

In this section, we briefly review some aspects of the low-energy effective
field theories for topological insulators and explain 
the relevance of the loop model in this context.

\subsubsection{Bulk topological theory in $(3+1)$ dimensions}

The topological insulators are characterized by time-reversal
($\cal T)$ symmetry \cite{fradkin-book}. Like other topological phases
of matter they possess a bulk gap and surface massless
excitations. These can have two field theory descriptions: i) in terms
of free massless fermions in the case of non-interacting (band)
systems \cite{bernevig-book} and ii) in terms of bosonic degrees of
freedom stemming from the bulk topological gauge theory
\cite{moore-BF}.  The bosonic approach is believed to be superior for
modeling interacting systems.

At energies below the bulk gap, the global effects
are accounted for by a topological theory.
In $(3+1)$ dimensions, this is given by the so-called BF theory \cite{moore-BF}:
\be
\label{S_BF}
S_{BF}[a,b,A] = i\int_{\mathcal{M}} {\dfrac{k}{2 \pi} b da +
  \dfrac{1}{2 \pi} b dA + \dfrac{\theta}{8 \pi^{2}} da da }\, .
\ee
The action involves the one and two-form hydrodynamic fields $a=a_\mu
dx^\mu$ and $b=(b_{\mu\nu}/2)dx^\mu dx^\nu$, that are dual to the
conserved currents for vortex-line and particle bulk excitations,
$V_{\mu\nu}$ and $J_\nu$, respectively: $V=*da$ and $J=*db$.
 The BF theory provides relative Aharonov-Bohm phases to these excitations. 
The coupling constant $k$ is a positive
integer, odd (even) for fermionic (bosonic) systems, the values $k=1$
being relative to free fermions and $k>1$ to interacting systems.

The BF action includes the background gauge field $A=A_\mu dx^\mu$ and
is ${\cal T}$ invariant only when the coupling $\th\sim\th +2\pi$ 
takes the values $\th=0$ or $\th=\pi$, the latter characterizing 
the nontrivial phase. By integrating
out the $a$ and $b$ fields, one obtains the induced action:
\be
S_{ind}[A] =  i\dfrac{\theta}{8 \pi^{2}k^2} \int_{\mathcal{M}} dA dA ,
\qquad \qquad \th=\pi.
\label{S_theta}
\ee
This theta term is consistent with Dirac quantization condition 
provided that the minimal electric charge of the system is \cite{fradkin-q0}:
\be
e_0 =\frac{1}{k}\, .
\label{min_e}
\ee
This fractional value also occurs in the Aharonov-Bohm phases between bulk 
excitations. 

The physical interesting manifolds $\mathcal{M}$ possess a boundary
with dynamical surface degrees of freedom, whose action
should be specified. Let us consider the expression:
\begin{equation} 
\label{S_surf}
S_{surf}[\zeta,a,A] =i\int_{\de \mathcal{M}} d^{3}x\left(
\dfrac{k}{2 \pi}   \varepsilon^{\mu \nu \rho} \zeta_{\mu} \de_{\nu} a_{\rho} +
\dfrac{1}{2 \pi} \varepsilon^{\mu \nu\rho} \zeta_{\mu} \de_{\nu} A_{\rho}
\right)\, ,
\end{equation}
involving the boundary values of the $a, A$ fields
and $\z=\z_\mu dx^\mu$, the restriction of $b$ to the
boundary \cite{moore-BF}.
This action is determined by the requirement of gauge invariance 
of the bulk-boundary system. Actually, the complete action $S_{BF}+S_{surf}$ is
invariant under $a\to a +d \l, b\to b+d\xi$ and $\z\to\z+\xi$.
 
 Note that the action (\ref{S_surf}) does not yet include any dynamics for the
surface degrees of freedom, because its Hamiltonian vanishes.
Introducing a dynamics by adding terms
to $S_{surf}$ will be the goal of the following discussion.
However, we should first discuss the boundary conditions for quantization.

\subsubsection{Solitonic modes}

The three-dimensional excitations of particles and vortex-lines are
sources for the $b$ and $a$ field equations of motion, respectively.
Placing such excitations in the bulk determines the boundary conditions
for the fields at the surface and thus introduce solitonic modes.

\begin{figure}
\includegraphics[scale=.5]{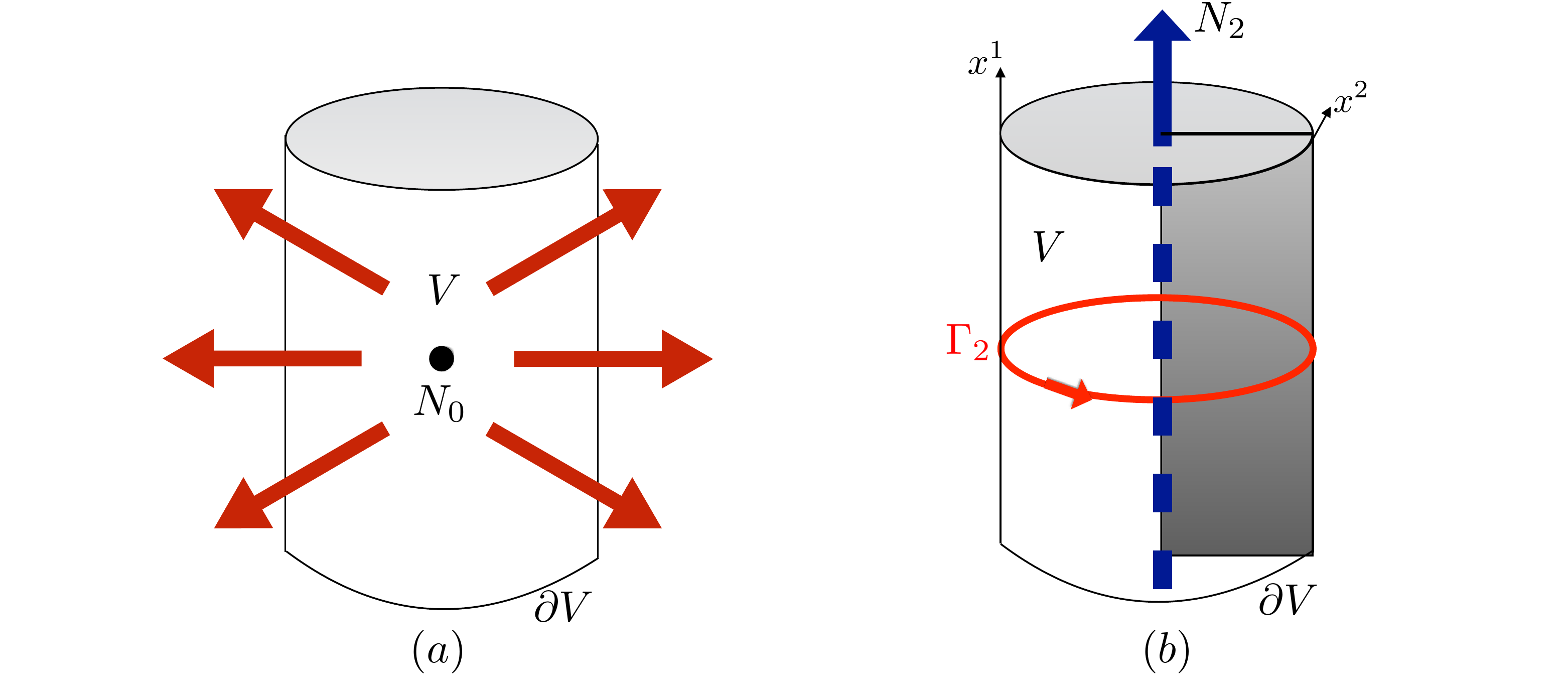}
\caption{Relation between bulk and boundary excitations:
(a) A charge $N_0$ in the bulk creates a flux for the $\z$
  field on the boundary.  (b) A vortex line with magnetic charge $N_2$
  along the cycle $\G_1$ gives a non-vanishing loop integral of the
  $a$ field along $\G_2$ on the surface.} 
\label{monod_t}
\end{figure}

Let us consider the spatial three-dimensional geometry of the solid
torus ${\cal M}=S_1\times S_1\times I$, whose boundary is the
two-torus ($I$ is the interval $[0,1]$).  
The possible bulk excitation are summarized in
Fig.\ref{monod_t}. In part (a), a static particle is put at
the origin: being the source for the $b$ field, it implies a non-vanishing
flux for the $\z$ field on the boundary torus.  
In part (b), a static vortex line winds along the cycle
$\G_1$ inside the solid torus: this is a source for the
$a$ field, whose line integral on the cycle $\G_2$ 
is non vanishing. Another condition exists by exchanging the two directions.

Therefore, the following 
boundary conditions are obtained for solitonic modes 
of the $\z$ and $a$ fields \cite{ryu-Z}:
\begin{align} 
&\int_{\mathbb{T}^{2}}{ d^{2}x \,\varepsilon^{ij}\de_{i} \zeta_{j} } =
  \dfrac{2 \pi N_{0}}{k}, \quad N_{0}\in \mathbb{Z}\,
  , \label{monod_z} \\ \nonumber 
\\ &\int_{\Gamma_{1}}{
    dx^i a_i} = \dfrac{2 \pi N_{1}}{k}\, , \quad\qquad N_{1} \in
  \mathbb{Z}\, , \label{monod_a1} 
\\ \nonumber
  \\ &\int_{\Gamma_{2}}{ dx^i a_i} = \dfrac{2 \pi N_{2}}{k}\, ,
  \quad\qquad N_{2} \in \mathbb{Z}\, . \label{monod_a2}
\end{align}

\subsubsection{Local surface dynamics}

In earlier works \cite{ryu-Z} \cite{magnoli-BF} \cite{CRS}, a simple dynamics for
surface excitations was introduced and studied. Let us review
some points of this analysis because they will be relevant for this
work.
The boundary action (\ref{S_surf}) for vanishing background, 
can be written in the static gauge $a_0=\z_0=0$ as follows\footnote{
In this Section, we use Minkowskian notation.},
\be
S_{surf}[a,\z,0]=\frac{k}{2\p}\int d^3x\, 
\varepsilon^{ij}\z_{i}\de_0 a_j.
\label{S_sympl}
\ee
This expression is the symplectic form for two pairs of canonically
conjugate degrees of freedom: the first one
is given by the longitudinal part $a_i=\de_i\vf$ and the
transverse part $\z_i^\perp$, $\de_i\z_i^\perp=0$; the second one
involves the transverse $a_i^\perp$ and longitudinal $\z_i=\de_i\l$
components. Disregarding the second pair, there remains the scalar field $\vf$
and its momentum $\Pi = (k /2 \pi) \eps^{ij}\de_{i} \zeta_{j}$.
The simplest dynamics is obtained by adding a quadratic Hamiltonian for
$\vf $, as follows:
\begin{equation} 
\label{S_surf_loc}
S_{surf}[\vf]=\int d^3x\, \Pi \de_0\vf -\mathcal{H}, \qquad
\mathcal{H} =  \dfrac{1}{2m} \Pi^{2} +\dfrac{m}{2} (\de_{i} \varphi)^{2}.
\end{equation}
The presence of the mass parameter $m$ is due to the mismatch
between the original dimension of gauge fields, implying a dimensionless $\vf$,
and the standard scalar field dimension.

The boundary theory \eqref{S_surf_loc} can 
be written in Lagrangian form: 
\be 
S_{surf}[\vf]=\dfrac{m}{2}\int d^3x\,\de_{\m}\varphi\de^\m\varphi.  
\label{S_surf_lag}
\ee 
Furthermore, the Hamilton equations in covariant form read,
\begin{equation} 
\label{S_surf_eom}
\dfrac{k}{2 \pi} \varepsilon^{\mu \nu \rho} \de_{\nu} \zeta_{\rho} = m
\de^{\mu} \varphi\, ,
\end{equation}
that reminds of the electric-magnetic duality in $(2+1)$ dimensions between a
gauge field $\z$ and a dual scalar $\varphi$ \cite{magnoli-BF}.

The quantization of the surface theory (\ref{S_surf_loc}) in presence
of the solitonic modes, Eqs.(\ref{monod_z}-\ref{monod_a2}),
and the properties of the spectrum were obtained in the works
\cite{ryu-Z} \cite{CRS}. However, this theory is not completely satisfactory
because it does not matches the free fermion dynamics in any limit.
Let us compute the induced action in presence of the $A$
background. The coupling of $\vf$ to $A$ is dictated by the bulk
theory and amounts to the substitution
$\de_\m\varphi\rightarrow\de_\m\varphi+A_\m/k$.  The action obtained
by integrating $\vf$ reads:
\begin{equation} 
\label{S_surf_ind}
S_{ind}^B[A] = \dfrac{m}{4 k^{2}} \int d^3x d^3y  \, F_{\mu \nu}(x)
  \left(\dfrac{1}{\de^2} \right)_{x,y} F^{\mu \nu} (y)\, ,
\end{equation}
using the notations introduced in Section 2.1.
Note that the mass $m$ appears explicitly, and cannot be eliminated by
a redefinition of the $\vf$ field, since its coupling to $A$ is fixed.

On the other hand, the fermionic induced action can be computed 
by expanding the determinant to leading quadratic 
order in $A$, corresponding to the semiclassical, weak-field approximation. 
One finds \cite{redlich}:
\be
\label{S_ind_F}
S_{ind}^{F}[A]=\dfrac{1}{64}\int \,
F_{\m\n}\dfrac{1}{\de}F^{\m\n} \pm
\frac{1}{8\pi}\int AdA\, .
\ee 
As explained in \cite{CRS}, the Chern-Simons term corresponding to the
parity anomaly is cancelled
by the bulk BF theory and should be disregarded.

We observe that the two expressions $S^B_{ind}$ and $S^F_{ind}$ differ
qualitatively in the low-energy limit: 
the fermion theory is conformal invariant and its induced action
does not include any mass scale; on the contrary, the bosonic
action contains the unavoidable mass $m$.
In conclusion, the local bosonic theory (\ref{S_surf_loc}) describe a solvable
surface dynamics that is different from that of 
topological band insulators. It may describe interacting fermions
in a spontaneously broken phase \cite{CRS}.


\subsubsection{Nonlocal surface dynamics and the loop model}

In the earlier work \cite{CRS}, it was argued that a nonlocal modification
of the action (\ref{S_surf_loc}) could bring closer to the fermionic
theory. Actually, the loop model provides 
the correct dynamics. Let us add its action (\ref{S_loop_ker}) with
couplings $(g,f)=(g_0,0)$ to the topological term
(\ref{S_surf}), as follows:
\begin{equation} 
\label{S_surf_loop}
S_{surf}[a,\, \zeta,\, A] = \dfrac{i}{2 \pi} \int{ \left(k \zeta da +
  \zeta dA \right)} + \dfrac{g_{0}}{4\pi} \int
  a_{\mu} \left( \dfrac{-\delta_{\mu \nu} \de^{2}+\de_{\mu}
    \de_{\nu}}{\de}\right) a_{\nu}\, .
\end{equation}
The integration of the field $\zeta$ implies the constraint $a=A/k$
and leads to the induced action,
\begin{equation}
	S_{ind}[A] = \dfrac{g_{0}}{4 \pi k^{2}} \int{d^{3}x d^{3}y\,
          \, A_{\mu} \left(\dfrac{-\delta_{\mu \nu}
            \de^{2}+\de_{\mu}\de_{\nu}}{\de}\right) A_{\nu}},
\label{S_ind_nl}
\end{equation}
which reproduces the expected fermionic result
\eqref{S_ind_F} for $k=1$ and $g_0=\pi /8$. 

Furthermore, the equation of motion for $a$ gives
a nonlocal generalization of the previously seen electric-magnetic duality 
\eqref{S_surf_eom},
\begin{equation} 
\label{em_d}
	-i \frac{k}{2\pi} \varepsilon_{\mu \nu \rho} \de_{\nu}
        \zeta_{\rho} = \dfrac{g_{0}}{2\pi}\de\, a_{\mu}\, ,
\end{equation}
that had been heuristically suggested in \cite{CRS} \cite{magnoli-BF}.

Therefore, the physics of topological insulators provides a strong
motivation for analyzing the loop model, as it represents a viable
theory for boson-fermion correspondence in the semiclassical,
weak-field limits.  The issue of bosonization and the meaning of the
quadratic approximation will become more clear in the following
sections.

Another form of the surface action is obtained by integrating out the
hydrodynamic field $a_\m$ in \eqref{S_surf_loop}. Upon using the
kernel identity (\ref{g_f_tilde}), we obtain:
\begin{equation} 
\label{S_surf_z}
S_{surf}[\zeta,A]=\dfrac{k^{2}}{4 \pi g_{0}} \int\, \zeta_{\mu}\biggr( 
\dfrac{-\delta_{\mu \nu} \de^{2}+\de_{\mu}\de_{\nu}}{\de}\biggr) \zeta_{\nu} +
\frac{i}{2\pi} \int\,   \zeta dA\, .
\end{equation}
This again corresponds to the loop model with its coupling to the
$A$ background and coupling constants:
\begin{equation} \label{g_0}
g = \dfrac{k^{2}}{g_{0}}, \qquad\quad f=0\, .
\end{equation}

\subsubsection{Partition function of the local theory}

The partition function of
the local theory (\ref{S_surf_loc}) on the three torus $\mathbb{T}^3$
was found in Refs.\cite{ryu-Z} \cite{CRS}. 
Let us recall its expression in the case of orthogonal axes,
of spatial radii $R_{1}$ and $R_{2}$ and time period $\beta$.
The canonical quantization of the $\z,a$ fields involves oscillator
and solitonic modes satisfying the conditions (\ref{monod_z}-\ref{monod_a2}).
The partition function correspondingly factorizes into
$Z=Z_{sol}Z_{osc}$. The first part reads:
\begin{equation} \label{Z_sol_loc}
Z_{sol} = \sum_{N_0,N_1,N_2 \in \mathbb{Z}}{\exp{\left\lbrace
    -\beta \left[ \dfrac{ N_{0}^{2}}{8 \pi^{2}R_{1}R_{2}m} + 2 \pi^{2}
      \dfrac{m}{k^{2}} \left(N_{1}^{2} \dfrac{R_{2}}{R_{1}} +
      N_{2}^{2} \dfrac{R_{1}}{R_{2}} \right)\right] \right\rbrace}}\, .
\end{equation}

The oscillator part $Z_{osc}$ is found by zeta-function regularization
of the determinant of the Euclidean Laplacian \cite{cc}.
The result takes the standard form of Bose statistics
times a Casimir energy term:
\ba
Z_{osc} &=& \left[\text{det}^\prime(-\de^2)\right]^{-1/2}
=e^{F} \prod_{(n_{1},n_{2}) \neq (0,0)}{\biggr(
  1-\exp{\biggr(-\dfrac{\beta}{2\pi}
    \sqrt{\dfrac{n_{1}^{2}}{R_{1}^{2}} + \dfrac{n_{2}^{2}}{R_{2}^{2}}}
    \, \, \biggr)} \biggr)^{-1}}\, ,
\nl& F& = \dfrac{\beta}{2R_{1}^{2}R_{2}^{2}} \sum_{(n_{1},n_{2})\neq
  (0,0)}\biggr[\dfrac{n_{1}^{2}}{R_{1}^{2}} +
  \dfrac{n_{2}^{2}}{R_{2}^{2}} \biggr]^{-3/2}\, ,
\label{Z_osc_t}
\ea
where the prime indicates the exclusion of zero modes.

The expression of the partition function will be useful for checking 
the quantization of the loop model described in Section 3.


\subsection{Duality relations in the loop model} 
  
Dualities indicate the possibility of representing a physical system
with two different field theories, say by using bosonic or fermionic
degrees of freedom.  In (2+1) dimensions, it is well-known that
non-relativistic particles can change their statistics by
coupling to a Chern-Simons gauge field.  Recently,
this mechanism was argued to hold for relativistic theories as well,
leading to several conjectures that fit into a ``web of
dualities'' \cite{dual-rev}. For instance, the fermion-boson duality 
reads \cite{tong}:
\begin{equation} \label{b_f_d}
	\mathcal{L}_{B}[\varphi] +
        J_{B\mu}a_{\mu}+\dfrac{i}{4\pi}ada+\dfrac{i}{2\pi}adA \sim
        \mathcal{L}_{F}[\psi]+ J_{F\mu}A_{\mu} -\dfrac{i}{8\pi}AdA\, .
\end{equation}

The theories on both sides of this relation are coupled to 
the external background field $A_\mu$. On the l.h.s, the bosonic current 
is first coupled to a dynamic Chern-Simons field
that changes the statistics from bosonic to fermionic by adding a
quantum of flux for each particle. On the r.h.s., the fermion 
parity anomaly term $(1/8\pi)AdA$ is subtracted.
 The duality relation is supposed to map not
only kinematical quantities such as spin and charge, but also the
low energy dynamics, even in the massless case.
For instance, the Abelian Higgs model at the critical
point is believed to be dual to a massless Dirac fermion \cite{dual-rev}.
The matter actions
$\mathcal{L}_{B}[\varphi] $ and $\mathcal{L}_{F}[\psi]$ 
include self-interactions suitably tuned for the duality to hold.

In this context, it is interesting to analyze the loop model, in which
the dualities are exact transformations, and are represented by
$SL(2,\Z)$ maps of the complex coupling constant $\t=f+ig$.

\subsubsection{Bosonic particle-vortex duality}

The bosonic particle-vortex duality is schematically written as follows
\cite{tong}:
\begin{equation} 
\label{b_b_d}
	\mathcal{L}_{B}[\phi]+j_{\mu}^{(\phi)}A_{\mu} \sim
        \widetilde{\mathcal{L}}_{B}[\varphi] +
        j_{\mu}^{(\varphi)}a_{\mu} +\dfrac{i}{2\pi}adA\, .
\end{equation}
In this expression, on the l.h.s. the charge density $j_0^{(\f)}$ of the
$\f$ field couples to the electric potential $A_0$: on the r.h.s., the
$a_0$ equation of motion imply $j_0^{(\vf)}\propto \eps^{ij}\de_i A_j$,
meaning that the dual bosonic field $\vf$ is magnetically charged.
This fact explains the name of particle-vortex, or electric-magnetic
transformation.

The partition functions of the two theories in the external
background, $Z[A]$ and $\wt{Z}[A]$, are related by the following map:
\begin{equation}
  \label{Z_d}
  Z[A] = \int \mathcal{D}a_{\mu} \, \widetilde{Z}[a]
    \exp{\left( \frac{i}{2\pi} \int{adA} \right)}\, .
\end{equation}
Let us compute this transformation for the loop model (\ref{S_loop_ker})
coupled the $A_\mu$ background, whose induced action can be found
by generalizing the derivation  of (\ref{S_ind_nl}) in Section 2.3.4: 
\be
Z[A]=\exp{\left(-\frac{1}{4\pi}\int A_\mu D_{\mu\nu}(g,f)A_\nu\right)}\, .
\label{S_ind_ker}
\ee
By performing the Gaussian integral in (\ref{Z_d}) and using the
kernel identity (\ref{g_f_tilde}), we obtain that $\wt{Z}[A]$ 
takes the same form
as $Z[A]$ with complex coupling constant:
\be
\wt\t = -\frac{1}{\t}\, ,\qquad\quad \t=f+ig\, ,
\label{s_d}
\ee
corresponding to the $S$ generator of the $SL(2,\Z)$ group.

Therefore, the loop model is explicitly self-dual \cite{fradkin-loop}.
The physical meaning of this result will be more clear in the
following, where we shall see that this theory corresponds to
electrodynamics in the limit of large number of matter fields $N$.

A nice aspect of the duality transformation (\ref{b_b_d}) is that
it actually corresponds to a Legendre transformation. 
Let us rewrite it,
\be
\wt{S}[{\cal J}] =S[A] -\int {\cal J}^\mu A_\mu, \qquad\quad {\cal J}= 
\frac{1}{2\pi} *\!(da),
\label{leg_d}
\ee
namely as a change of variable 
from the background $A$ to the ``effective field'' ${\cal J}$, where
the new ``effective potential'' $\wt{S}[{\cal J}]\equiv\wt{S}[a]$ is 
equal to the dual action. As is well known, the second derivatives
of the two potentials $S$ and $\wt{S}$ w.r.t. the respective variables are one
the inverse of the other, 
$\d^2 S/\d A_1 \d A_2 \sim 
\left(\d^2 \wt{S}/\d {\cal J}_1\d{\cal J}_2\right)^{-1}$.
The first variation w.r.t. to the background defines the induced current,
while the second derivative introduces the conductivity. As a consequence,
the duality implies a reciprocal relation between conductivity
tensors, $\s_{ij}(\t)$ and $\wt{\s}_{kn}(\wt{\t})$, $i,j,k,n=1,2$, 
of the two theories, as follows \cite{son-F}:
\be
\eps^{ij}\,\s_{jk}(\t)\,\eps^{kn}\,\wt{\s}_{nm}(\wt{\t})= 
\frac{1}{4\pi^2}\d^i_m \, .
\label{cond_d}
\ee


\subsubsection{Fermionic particle-vortex duality}

The electric-magnetic duality for fermionic theories is conjectured
to take form \cite{tong}:
\begin{equation} 
\label{f_f_d}
\mathcal{L}_{F}[\psi]+j^{(\psi)}_{\mu}A_{\mu} 
\sim \widetilde{\mathcal{L}}_{F}[\chi]+ j^{(\chi)}_{\mu} a_{\mu} +
\dfrac{i}{4\pi} adA\, ,
\end{equation}
between the fermion field $\psi$ and its dual $\chi$.
The map is the same as for bosonic fields (\ref{b_b_d}) up to a 
normalization of the statistical field $a$.

As it will be clear in the following, the loop model describes
both (the large $N$ limit of) bosonic and fermionic theories; thus,
we can apply the map (\ref{f_f_d}) to the effective action 
(\ref{S_ind_ker}) again
and obtain the relation (\ref{s_d}) between the couplings up to
a factor of four.  Upon defining the
``fermionic'' version of the loop model with shifted coupling 
$\t_F=2\t$, we can write the fermionic duality (\ref{f_f_d}) as:
\be
\widetilde{\t}_F=-\frac{1}{\t_F}, \qquad\quad \t_F=2\t\equiv 2\t_B .
\label{s_f_d}
\ee

\subsubsection{Boson-fermion duality}

Let us now consider the transformation in Eq.\eqref{b_f_d}:
on the bosonic side, first a Chern-Simons term $ada$ is added
and then the particle-vortex transformation (\ref{b_b_d}) is applied.
Acting on the loop model, these correspond to
the following maps:
\be
T: \t_B\ \to\ \t_B+1, \qquad\qquad S: \t_B+1\ \to \ -\frac{1}{\t_B+1}\, .
\label{b_f_d2}
\ee
On the fermionic side, the subtraction of the anomaly term corresponds
to $T^{-1}: \t_F\to \t_F -1$, taking into account the different normalization
of the fermionic model (\ref{s_f_d}).
In conclusion, the combined map is:
\be
-\frac{1}{\t_B+1}=\frac{\t_F-1}{2}\qquad \longrightarrow\qquad
\t_F=\frac{\t_B-1}{\t_B+1}\, .
\label{b_f_d3}
\ee
Therefore, the loop model explicitly realizes the boson-fermion duality too.

In the literature, the dualities of Abelian theories in $(2+1)$ dimensions
have been related to those of Yang-Mills theory in $(3+1)$ dimensions
\cite{witten} \cite{dual}.
This can be easily explained within the bulk-boundary
correspondence discussed in Section 2.3: the topological
bulk action (\ref{S_BF}) possesses the theta-term $\th/8\pi^2\int dada$, that
under periodicity, $\th \to \th+2\pi$, produces a Chern-Simons action
at the boundary corresponding to the $T$ transformation $\t_B\to\t_B+1$
discussed above. Therefore, the dualities involving bosonic theories 
include the transformations $T$ and $S$ that span the
$SL(2,\Z)$ group \cite{cft}.

On the other hand, the dualities within fermionic theories also belong
to the $SL(2,\Z)$ group: the transformation $S$ was found in
(\ref{s_f_d}), while $T: \t_F \to \t_F+1$ is obtained by integrating
out one fermionic degree of freedom as in (\ref{S_ind_F}).

The boson-fermion map (\ref{b_f_d3}) can be written
group theoretically as follows: 
\be
\t_F=T\L ST (\t_B)\, .
\label{b_f_gr}
\ee
There appears another transformation $\L: \t_B \to \t_F=2\t_B$  
that does not belong to the $SL(2,\Z)$ group:
in matrix notation, this is diagonal, $\L={\rm diag}(2,1)$, with
determinant two.
However, this transformation cannot be iterated, i.e. $\L^n$ does
not make sense for $n\in \Z$, beside $n=0,\pm 1$. 
Thus, it is not an ordinary group element and does not enlarge
the duality group.

In conclusion, dualities including both bosonic and
fermionic theories belong to the group $SL(2,\Z)$, keeping in mind 
the coupling normalization just discussed.
In the following, we do not discuss these issues any further because
we are mostly concerned with the analysis of the bosonic loop model
with vanishing Chern-Simons term ($f=0$), for which the inversion
$g\to 1/g$ suffices.


\subsection{Electrodynamics in the large-$N$ limit and loop model}

In this section, we discuss the theories of $(2+1)$-dimensional particles
(both fermionic and bosonic) interacting with photons in $(2+1)$
and $(3+1)$ dimensions, corresponding to $QED_{3}$, and its
mixed-dimensional modification $QED_{4,3}$ \cite{son-F} \cite{son-B}.
We show that they reduce to the loop model in the limit of large number
of matter fields.

\subsubsection{Loop model and $QED_{3}$}

The action of $QED_3$ with $N_F$ massless fermionic fields is,
\begin{equation} 
\label{S_QED3}
S_{QED_{3}}[\psi,A] =\int{d^{3}x\, \sum_{n=1}^{N_{F}} \bar{\psi}_{n}(i
  \slashed{\de} -\slashed{A})\psi_{n}} + \dfrac{1}{4e^{2}}
\int{d^{3}x\, F_{\mu \nu}F_{\mu \nu}}\, .
\end{equation}
Integration of the fermions produces the determinant of the Dirac operator
raised to the $N_F$ power: a simplification occurs in the large $N_F$-limit
by keeping the coupling $\l=e^2 N_F$ finite, because the expansion
of the determinant in powers of $A_\mu$ is dominated by
the quadratic term, the higher orders
being subdominant by powers of $N_F^{-1/2}$.
The expression of the quadratic term is equal to the induced action
already given in Eq.(\ref{S_ind_F}): thus, the large-$N_F$ limit is,
\ba
Z&=& \int {\cal D}A \, \exp\left(-S_{QED_3}[A] \right),
\nl
\!\!\!\!
S_{QED_{3}}[A]&= &\dfrac{\l}{2}\int
    A_{\mu}\left(\dfrac{1}{16}\dfrac{1}{\de}+\dfrac{1}{\lambda}\right)
    \left(-\delta_{\mu \nu}\de^{2}+\de_{\mu}\de_{\nu}\right)A_{\nu}
    \, + i\frac{\eta}{8\pi}\int AdA.
\label{S_QED3_N}
\ea 
The parity anomaly term has a $\pm$ sign ambiguity for each
fermion component, that can be resolved by considering the 
limit $m\to 0^\pm$ of massive fields \cite{redlich}. Without knowing this
information or other physical input on the theory, we can only say
that the parameter $\eta$ in (\ref{S_QED3_N}) is an integer taking one
value in the interval $-N_F\le \eta \le N_F$.

Next we observe that in the first part of the action (\ref{S_QED3_N}), 
the term $1/\l$ involves a mass scale and is
subdominant w.r.t. $1/\sqrt{k^2}$ in the low-energy limit.
We conclude that the effective large-$N_F$/low-energy theory of $QED_{3}$
is described by the loop model for values of the couplings
$(g,f)=(\p /4,\eta/\l)$ (using the fermion normalization (\ref{s_f_d}) and
after rescaling the field $A_\mu\to A_\mu/\sqrt{\l}$).


\subsubsection{Loop model and $QED_{4,3}$} 
\label{mixed_QED}

The action of this model \cite{son-F},
\begin{align} 
\label{S_QED43}
S_{QED_{4,3}}[\psi,\, A]&= \int_{{\cal M}_3}
{d^{3}x\, \sum_{n=1}^{N_{F}}\bar{\psi}
  (i\slashed{\de} -\slashed{A}) \psi} +
\dfrac{1}{4e^{2}}\int_{{\cal M}_4}{d^{4}x\,
  F_{\mu \nu}F_{\mu \nu}}\, ,
\end{align}
shows that the photons are defined in $(3+1)$ dimensions while
the fermions live on a $(2+1)$-dimensional hyperplane.
This theory is very interesting because it maps into itself
under the fermionic particle-vortex duality \eqref{f_f_d}
\cite{son-F}. Let us review this result for $N_F=1$.

The integration of the $A_\mu$ field in (\ref{S_QED43}) leads to the
term $\int j^{(3)}_\mu (x)D^{(4)}_{\mu\nu} (x-y) j^{(3)}_\nu(y)$, where the
three-dimensional currents interact with the four-dimensional 
propagator restricted to the hyperplane.
We denote the coordinates as $X^\mu=(x^\a,x^3)$, and identify
the hyperplane by $x^3=0$.
The Green function of the four-dimensional 
Euclidean Laplacian $\de^2_{(4)}$ on the hyperplane can be written as:
\be
\left.\frac{1}{-\de^2_{(4)}}(X,Y)\right\vert_{x^3=y^3=0}=
\left.\frac{1}{4\pi^2}\frac{1}{(X-Y)^2}\right\vert_{x^3=y^3=0}=
\frac{1}{2}\frac{1}{\de}(x,y) \, ,
\label{green_34}
\ee
i.e. it corresponds to the kernel of the loop model.
Therefore, the integration of the gauge field leads to the following
three-dimensional action with long-range current-current interaction:
\begin{equation} 
S_{QED_{4,3}}[\psi] = \int_{{\cal M}_3} \bar{\psi}i\slashed{\de} \psi 
+ \dfrac{e^{2}}{4}\  j^{(\psi)}_{\mu}\,
  \dfrac{1}{\de}\, j^{(\psi)}_{\mu} \, ,
\label{S_jj}
\end{equation}
with $j_{\mu}^{(\psi)} = \bar{\psi} \gamma_{\mu} \psi$. 

The dual theory with coupling constant $\wt{e}$ is obtained by applying
the particle-vortex transformation (\ref{f_f_d}) to (\ref{S_QED43}):
\begin{align} 
\label{S_QED43_d}
\widetilde{S}_{QED_{4,3}}[\chi,a,A] &= \int_{{\cal M}_3}{ d^{3}x\, \left[
    \bar{\chi}(i \slashed{\de} -\slashed{a}) \chi -\dfrac{i}{4\pi}
    \varepsilon_{\mu \nu \rho} a_{\mu} \de_{\nu} A_{\rho} \right]}
\nonumber \\ & +\dfrac{1}{4\wt{e}^{2}} \int_{{\cal M}_4}{d^4x\, F_{\mu \nu}F_{\mu
    \nu}}\, ,
\end{align}
where $a_\mu$ is the statistical field.
Integration over the $A_\mu$ field following the same steps as before 
leads to the three-dimensional action:
\be 
\label{S_QED43_d2}
\wt{S}_{QED_{4,3}}[\chi,a] = \int_{{\cal M}_3} \bar{\chi}(i
  \slashed{\de} -\slashed{a}) \chi+
\dfrac{\wt{e}^{2}}{64\pi^2} \ a_{\mu}
  \biggr( \dfrac{-\delta_{\mu \nu} \de^{2}+\de_{\mu} \de_{\nu}}{\de}
  \biggr) a_{\nu}\, .
\ee
Finally, integrating out $a$ with the help of the loop-model identity
(\ref{g_f_tilde}) gives,
\begin{equation} 
\label{S_jj_d}
\widetilde{S}_{QED_{4,3}}[\chi] = \int_{{\cal M}_3}
  \bar{\chi}i\slashed{\de} \chi  +
\dfrac{16\pi^{2}}{\wt{e}^{2}}\ j^{(\chi)}_{\mu}\,
  \dfrac{1}{\de}\, j^{(\chi)}_{\mu}  \, ,
\end{equation} 
where $j^{(\chi)}_{\mu}=\bar{\chi} \gamma_{\mu} \chi\,$.

The comparison of the actions \eqref{S_jj} and \eqref{S_jj_d}
establishes the self-duality of $QED_{4,3}$ with coupling constant
relation:
\begin{equation} 
\label{e_d} 
\wt{e} = \dfrac{8 \pi}{e} \, .
\end{equation} 
The duality implies a inverse relation between the conductivities of the
two theories, as discussed in Section 2.4.1 \cite{son-F}.
The same results is obtained in the case of electrodynamics of 
scalar particles \cite{son-B}; there is a 
difference of a factor of two in the relation (\ref{e_d}),
i.e. $\pi \to \pi/2$, stemming from the duality 
transformations (\ref{b_b_d}) and (\ref{f_f_d}).

Let us now discuss the large $N_F$-limit of $QED_{4,3}$. 
It is convenient to start from the dual action
\eqref{S_QED43_d2}: the integration over the fermions
yields again the $N_F$ power of the determinant and its quadratic
approximation holds for $N_F\to\infty$  and $\l=e^2 N_F$ fixed, as in
the case of $QED_3$. We obtain the action:
\begin{equation} \label{S_QED43_largeN}
S_{QED_{4,3}}[a] = \biggr( \dfrac{1}{32}+\dfrac{1}{\lambda} \biggr)
\int\, a_{\mu} \biggr( \dfrac{-\delta_{\mu
      \nu}\de^{2}+\de_{\mu}\de_{\nu}}{\de} \biggr)a_{\nu}\, + \,
i\frac{\eta}{8\pi\l}\int a da \, ,
\end{equation} 
after rescaling of $a_{\mu} \to a_{\mu}/\sqrt{\l}$.

We conclude that $QED_{4,3}$ in the large $N_F$-limit is equivalent to
the fermionic loop model (\ref{S_loop_ker}) with coupling constant:
\begin{equation} 
\label{g_QED43}
g = \pi \left( \dfrac{1}{4} +\dfrac{8}{\lambda} \right)\, .
\end{equation}

This result is very important because it establishes that the loop model
is the limit of a viable theory of interacting electrons: for example, on the
surface of three-dimensional topological insulators discussed in Section 2.3,
the higher-dimensional photons can be physical and not merely a 
technical advantage.
In the next Section we shall see that the relation with $QED_{4,3}$ 
also provides a physical approach to quantize the loop model.

We conclude this section by adding some remarks:
\begin{itemize}
\item
Eq. (\ref{g_QED43}) shows that the dimensionless coupling constant $\l>0$ 
of $QED_{4,3}$ remap the critical line $g>1$ of the loop model.
Note that $QED_3$ is found at the point $\l=\infty$ on this line.
\item
It is believed that $QED_{4,3}$ possesses a critical line
also for finite $N_F$ \cite{beta-f}, that then spans $e^2<8\pi$ owing to 
the self-duality (\ref{e_d}). 
Note, however, that the finite-$N_F$ self-duality does not
survive the large $N_F$ limit and is replaced by the loop model
duality at $N_F=\infty$. 
\item
Finally, the analysis of scalar $QED_{4,3}$ in the large $N_B$
limit reproduces again the loop model up to numerical factors in the
coupling constant relation (\ref{g_QED43}). Indeed, the quadratic
expansion of the bosonic determinant has the same expression of the
fermionic theory, but without the anomalous Chern-Simons term.
\end{itemize}


\section{Quantization of the loop model on 
$\mathbb{T}^{3}$} \label{sec_Z_torus}

In this section we analyze the surface excitations of 
topological insulators with loop-model dynamics, 
as discussed in Section 2.3.4. We recall the expression of the action
(\ref{S_surf_loop}):
\begin{equation} 
\label{S_loop_quant}
S_{surf}[a,\, \zeta,\,0] = \dfrac{ik}{2 \pi}\int  \zeta da +
  \dfrac{g_{0}}{4\pi} \int  
a_{\mu} \left( \dfrac{-\delta_{\mu \nu} \de^{2}+\de_{\mu}
    \de_{\nu}}{\de}\right) a_{\nu}\, ,
\end{equation}
where the $A$ background has been switched off and
the anomalous Chern-Simons term is
cancelled by the bulk, so as to respect time-reversal symmetry
(coupling $f=0)$.
We consider the bulk geometry of the solid torus $\mathbb{T}^{3}\times I$.
The nontrivial part of the surface dynamics is given by the solitonic
excitations that are defined by the boundary conditions of the $\z$
and $a$ fields in (\ref{monod_z}-\ref{monod_a2}), 
corresponding to global magnetic and electric
fluxes on the spatial torus $\mathbb{T}^2$.

In addition, the compactness of the $a$ field allows for further
magnetic solitons. We place ourselves in the
massless phase of the loop model where local monopoles are suppressed
but global fluxes are possible on compact geometries. The corresponding
condition reads:
\begin{equation}
\int_{\mathbb{T}^{2}}{\varepsilon^{ij}\de_{i}a_{j}} = \dfrac{2 \pi
}{q_{0}}M_{0}, \qquad\qquad M_{0} \in \mathbb{Z}\, , 
\label{monod_a}
\end{equation}
where $q_0$ is the minimal charge for the $a$ field.

The usual method of quantization is based on expanding the
fields in solitonic and oscillator parts and evaluate the
partition function in terms of classical action and 
fluctuations around it. This analysis is not
possible for the nonlocal theory \eqref{S_loop_quant} that does not have
a Hamiltonian formulation and is not well defined on-shell. 

This problem can be solved by reformulating the loop model as a local
theory in $(3+1)$ dimensions, as we now explain.  We take some
inspiration from the mixed-dimension $QED_{4,3}$, where photons live
in $(3+1)$ dimensions and are coupled to a current confined to a
$(2+1)$-dimensional hyperplane. As seen in the previous section,
integration of the photons yields the nonlocal loop model interaction
on the surface.

\begin{figure} 
\begin{center}
\includegraphics[scale=1.5, trim={0 1.3cm 0 0}]{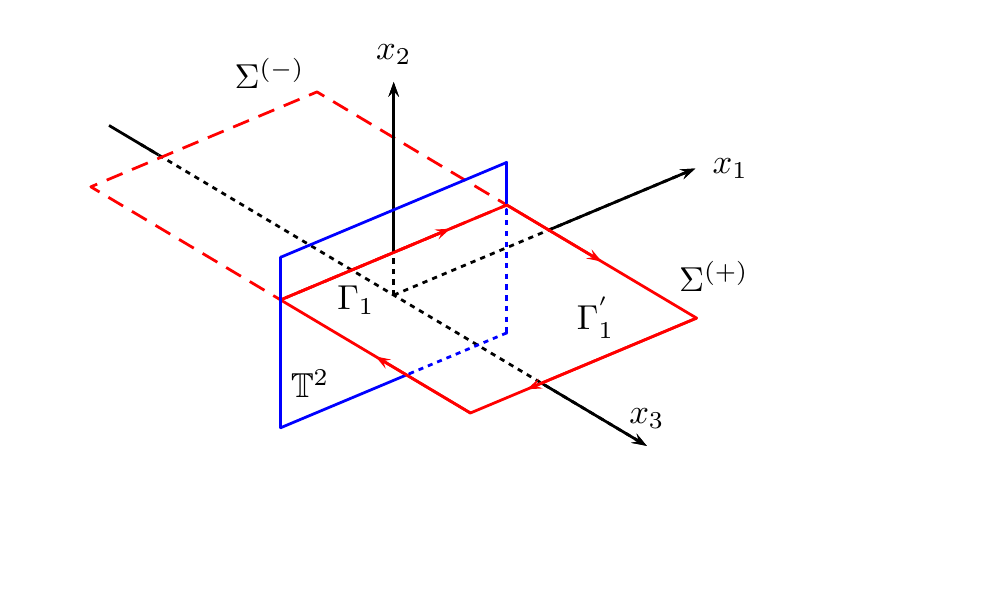}
\caption{Three-dimensional extension of the spatial torus
  $\mathbb{T}^{2}$. The torus is drawn in blue, while the surfaces
  $\Sigma^{(+)}$ and $\Sigma^{(-)}$ are in red.}
\label{fig_ext}
\end{center}
\end{figure}

We introduce an extra dimension and define the following action:
\begin{equation} 
\label{S_loop_4d}
	S_{4}[\hat{a},\zeta] = \dfrac{1}{4e^{2}}
        \int_{\mathcal{M}_{4}}d^{4}x\, \left(
\de_\mu\hat{a}_{\nu}-\de_\nu\hat{a}_{\mu} \right)^2
           +i \dfrac{k}{2\pi} \int_{\mathcal{M}_{3}}{ a
          d\zeta}\, .
\end{equation}
In this expression, the four-dimensional manifold is 
$\mathcal{M}_4=\mathbb{T}^3\times\mathbb{ R}$ with
extra coordinate $x_3\in\mathbb{ R}$,
$\hat{a}_\mu$ is the four-dimensional extension of the field
$a_\m$ on $\mathcal{M}_3= \mathbb{T}^{3}$ and 
$e$ is a coupling constant to be determined later.
The three-dimensional part of the action (\ref{S_loop_4d}) can be written
as a source term,
\begin{equation}
i\int_{\mathcal{M}_4} J^\mu  \hat{a}_{\mu}\, ,
\qquad\qquad 
J_\a= \delta(\mathcal{M}_{3})\dfrac{k}{2\pi} 
   \eps_{\a\b\g} \de_\b   \zeta_\g\, ,\qquad J_3=0,
\label{source_4d}
\end{equation}
where $\delta{(\mathcal{M}_{3})}=\delta(x_{3})$ is the delta function on the
hyperplane. The spatial part of this geometry is
drawn in Fig.\ref{fig_ext}.

The $(3+1)$-dimensional action (\ref{S_loop_4d}) corresponds to
ordinary electrodynamics that is well defined on-shell. We can compute
its partition function by decomposing the fields $\hat a$ and $\z$
into solitonic and oscillator parts:
\begin{equation} 
\label{Z_factor}
	Z = \sum_{sol\ config}
          e^{-S_{4}[\hat{a}_{sol}, \zeta_{sol}]}
        \int{\mathcal{D}\hat{a}_{osc}\, \mathcal{D}\zeta_{osc}
          e^{-S_{4}[\hat{a}_{osc},\zeta_{osc}]}}\, ,
\end{equation}
where $\hat{a}_{sol}$ are classical solutions of the $(3+1)$-dimensional
equations of motion obeying the $(2+1)$-dimensional 
boundary conditions for the $a,\z$
fields (\ref{monod_z}-\ref{monod_a2}) and (\ref{monod_a}).

Next the integration of wave modes of the field $\hat{a}$
in $S_4[\hat{a}_{osc},\zeta_{osc}]$, following usual steps, leads
to the $(2+1)$-dimensional action for (the wave modes of) $\z_\m$:
\begin{equation} 
\label{S_loop_z2}
S[\zeta] = \dfrac{k^{2}e^2}{16 \pi^2} \int
  \zeta_{\mu}\biggr( \dfrac{-\delta_{\mu \nu} \de^{2}
    +\de_{\mu}\de_{\nu}}{\de}\biggr) \zeta_{\nu}\, .
\end{equation}
This expression is the same as the that of the original surface action
$S_{surf}[a,\z]$ (\ref{S_surf_loop}), after eliminating the $a$ field
(cf. Section 2.3.4, Eq.(\ref{S_surf_z})), leading to the coupling
identification:
\be
\label{e_g}
e^2=\frac{4\pi}{g_0}.
\ee

In conclusion, the loop model (\ref{S_surf_loop}) has been transformed
into the local theory in $(3+1)$ dimensions (\ref{S_loop_4d}), that
allows for a proper definition and calculation of solitonic modes.

\subsection{Evaluation of solitonic modes} 
\label{torus_sol}

The $(3+1)$-dimensional Minkowskian action for static solitonic configuration
$S_4[\hat{a}_{sol}, \zeta_{sol}]$ corresponds to the Hamiltonian,
\begin{equation}
S_4[\hat{a}_{sol}, \zeta_{sol}]=-\b\, H,
\qquad	H = \dfrac{1}{2e^{2}} \int{d^{3}x \left(
          \bs{B}^{2}+\bs{E}^{2} \right)}\, ,
\label{H_torus}
\end{equation}
involving  the electric and magnetic fields $\bs{E}$
and $\bs{B}$ of $\hat{a}_\m$. 
The integration is done on the spatial part of $\mathcal{M}_4$ 
(cf. Fig.\ref{fig_ext}), specified by the torus periods,
$\bs{\omega}_{1}=(0,2\pi R_{1}, 0)$ and $\bs{\omega}_{2}=(0,0,2\pi R_2)$,
and by a finite interval, $x_3\in[-1/(2M), 1/(2M)]$, for the extra coordinate,
where $1/M$ is the infrared cutoff  to be discussed later.

Let us now solve the $\hat{a}_\m$ equations of motion with source
term (\ref{source_4d}). The magnetic flux configuration for $\z_\m$
on $\mathbb{T}^2$ \eqref{monod_z} determines a constant current $J_0$
on the $x_3=0 $ plane, which is coupled to $\hat{a}_0$ by the Poisson
equation:
\be
\nabla^2\hat{a}_0=-e^2J_0\, ,\qquad J_0=\d(x_3)\frac{N_0}{V^{(2)}},
\qquad V^{(2)}=4\pi^2 R_1 R_2\, .
\label{pois_torus}
\ee
The solution $\hat{a}_{0}=\hat{a}_{0}(x_{3})$ is easily found and it
determines the electric field component along $x_3$:
\begin{equation} 
\label{el_3}
	E_{3} = -\dfrac{d}{dx_{3}} \hat{a}_{0} =
        \dfrac{e^{2}N_{0}}{2V^{(2)}} \text{sign}(x_{3})\, .
\end{equation}
The contribution of the electric field to the Hamiltonian (\ref{H_torus})
is obtained by integrating over three-space, with the result:
\begin{equation}
	H_{el} =  \dfrac{e^{2}N_{0}^{2}}{32 \pi^2 R_1R_2 M}\, . 
\label{H_el} 
\end{equation}

Next we consider the configurations of electric flux for $a$ on
the $x^3=0 $ plane \eqref{monod_a1},
given by the line integral on $\Gamma_{1}$.
This can be extended to a close circuit
on the edge of the surface $\Sigma^{(+)}$ (cf. Fig.\ref{fig_ext});
two sides of this contour cancel each other
and the contribution $\G_1'$ at large $x_3$ vanishes by assumption.
Thus, the $a$ line integral can be rewritten as the flux  of the magnetic
field $B_2$ through $\Sigma^{(+)}$, leading to:
\begin{equation} 
\label{B_2}
	B_{2} =\dfrac{2M}{k} \dfrac{N_{1}}{R_{1}} \text{sign}(x_{3})\, .
\end{equation} 
In this expression, the sign function appears for the possible exchange
of $\Sigma^{(+)}$ with $\Sigma^{(-)}$.
In analogous fashion, the other flux condition (\ref{monod_a2})
determines a magnetic field along $x_1$:
\begin{equation} 
\label{B_1}
	B_{1} = \dfrac{2M}{k} \dfrac{N_{2}}{R_{2}} \text{sign}(x_{3})\, .
\end{equation}

Finally, the magnetic flux configuration for $a$ (\ref{monod_a})
on the $x^3=0 $ plane is reproduced by the following
$x_3$-independent field component:
\begin{equation} 
\label{B_3} 
B_{3} =   \dfrac{ M_{0}}{q_{0} 2\pi R_1 R_2}\, .
\end{equation}

The total magnetic contribution to the energy is then found to be:
\begin{equation} 
\label{H_mag}
  H_{mag} =\frac{M_0^2}{4\pi e^2 q_0^2 M R_1R_2}+
  \dfrac{8 \pi^{2}M}{e^{2}k^2}
  \left(N_{1}^{2} \dfrac{R_{2}}{R_{1}} + N_{2}^{2}\dfrac{R_{1}}{R_{2}}\right)
\, .
\end{equation}

From the evaluation of the classical solutions we thus obtain
the following expression of the solitonic part of the 
partition function of the loop model on $\mathbb{T}^3$:
\be 
\label{Z_sol_t} 
\!\! Z_{sol} = \!\! \sum_{N_\mu,M_{0}\in\mathbb{Z}}
\exp\left\{ -\beta \left[ \dfrac{2\pi g_0 M}{k^2}
\left(N_{1}^{2} \dfrac{R_{2}}{R_{1}} + N_{2}^{2}\dfrac{R_{1}}{R_{2}}\right)
+ \frac{1}{R_1R_2M}
\left(\frac{N_0^2}{8\pi g_0}+\frac{M_0^2g_0}{16\pi^2q_0^2}\right)\right]
\right\},
\ee
where we substituted the coupling $g_0$ using (\ref{e_g}).
Let us complete the calculation of the partition function before discussing
this result.

\subsection{Oscillator modes}

The partition function of the oscillator modes
can be obtained from the nonlocal $(2+1)$-dimensional 
Lagrangian (\ref{S_loop_z2}) by computing the 
determinant of the positive definite Euclidean Laplacian. 
Choosing the Lorentz gauge, the spectral decomposition reads:
\begin{equation} \label{S_osc}
S = \dfrac{k^{2}}{4 \pi g_{0}} \sum_{n_\a \in \mathbb{Z}^{3}\neq (0,0,0)} 
\zeta_\mu(n) \sqrt{\left(k_n\right)^{2}}
\zeta_\nu (n)\, ,
\end{equation}
where $k^\mu_n$ are the discretized momenta on $\mathbb{T}^3$. 
The field $\zeta$ possesses two physical polarizations,
instead of one for local Yang-Mills theory.
Thus, the oscillator part of the partition
function is given by the determinant,
\begin{equation} 
\label{Z_osc}
Z_{osc}= \left[ \text{det}^{\prime} \left( \sqrt{-\de^2}
  \right) \right]^{-1} 
= \left[\text{det}^{\prime} \left(-\de^{2}\right)\right]^{-1/2}\, .
\end{equation} 

As a matter of fact, this oscillator partition function is equal to that
of the local bosonic theory \eqref{S_surf_lag}, discussed in
Section 2.3.5.
We remark that $Z_{osc}$ is independent of the coupling constant.

In conclusion, the partition function of the loop model on $\mathbb{T}^3$
is given by $Z=Z_{sol}Z_{osc}$, where the expressions of 
$Z_{sol}$ and $Z_{osc}$ are given in (\ref{Z_sol_t}) and (\ref{Z_osc_t}), 
respectively.

\subsection{Interpolating theory and the choice of infrared cut-off}

The torus partition function found in the previous 
section possesses striking similarities with the corresponding quantity
in the local scalar theory of surface excitations discussed in Section 2.3.3.
The oscillator part take the same form;
regarding the solitonic sum, let us compare the expression
(\ref{Z_sol_t}) with the analogous one of the scalar theory
(\ref{Z_sol_loc}), reported in Section 2.3.5.  
We see that the terms parameterized by $N_0,N_1,N_2$ remarkably
match in the two formulas, 
upon identifying the respective mass parameters
by $m=Mg_0/\pi$. On the other hand, the $M_0$ term for
$a $ magnetic solitons is absent in the scalar theory, 
because the latter corresponds to the longitudinal part of the gauge field, 
$a_\mu=\de_\mu\vf$ (cf. Section 2.3.3)

This remarkable equivalence can be explained as follows: the two
theories are different, but can be matched on-shell.  For example, the
off-shell induced actions $S_{ind}[A]$ (\ref{S_surf_ind}) and
(\ref{S_ind_nl}) are unequal, and this fact originally motivated the
study of the nonlocal theory.

In order to understand these results, we reformulate the loop model by
introducing the infrared cutoff as an explicit photon mass
$\wt{M}$. The modification of the action (\ref{S_loop_quant}) reads:
\be
S_{m}=i\frac{k}{2\pi}\int \z da +\frac{g_0}{4\pi}
\int a_\mu\left(\frac{\d_{\mu\nu}(-\de^2+\wt{M}^2)+\de_\mu\de_\nu}
{\sqrt{-\de^2+\wt{M}^2}}\right)a_\nu \, .
\label{S_m}
\ee
In the Lorentz gauge, this becomes:
\begin{equation}
S_{m}[a,\zeta] = \dfrac{ik}{2 \pi} \int{ \zeta da }
+\dfrac{g_{0}}{4\pi}\int{d^{3}x\,d^{3}y\, a_{\mu}
  \sqrt{-\de^{2}+\wt{M}^{2}}\, a_{\mu}}\, .
\label{S_m_gauge}
\end{equation}
Upon integrating on $a$, this action describes conserved currents with
cutoffed long-range interaction:
$\int J_\mu(1/\de)J_\mu \to \int J_\mu(1/\sqrt{-\de^2 + \wt{M}^2})J_\mu$.
Therefore, $S_m$ can be considered as an equivalent formulation
of the loop model, where the cutoff is explicit and not added
a-posteriori in the classical field solutions.

Let us now analyze the theory on-shell: the equations of motion for $a$,
\begin{equation}
-i\dfrac{k}{2\pi} \varepsilon_{\mu \nu \rho} \de_{\nu}
   \zeta_{\rho} = \dfrac{g_{0}}{2 \pi} \sqrt{-\de^{2}+\wt{M}^{2}}
    a_{\mu} \rightarrow
\begin{cases}
& \dfrac{g_{0}}{2\pi} \de a_{\mu}\, , \quad \text{UV}\, , \\ \\
& \dfrac{g_{0}}{2\pi} \wt{M} a_{\mu}\, , \quad \text{IR}\, ,
\end{cases}
\label{eq_m_inter}
\end{equation}
interpolate between those of the nonlocal and local theories,
Eqs. (\ref{S_surf_eom}) and \eqref{em_d}. The equation of motion for
$\zeta_\m$ imposes $a_\m=\de_\m \varphi$: substituting in $S_{m}$, we
find the reduced action,
\begin{equation}
	S_{m} = \dfrac{\wt{M} g_{0}}{4\pi} \int{d^{3}x \, \de_{\mu}
          \varphi \de_{\mu} \varphi}\, , \qquad\qquad \text{(on-shell)}\, .
\label{S_m_shell}
\end{equation}
Therefore, the massive nonlocal action  $S_m$ (\ref{S_m}) is equal to the
local action \eqref{S_surf_lag} on-shell (up to a numerical factor).
This implies that the two theories have same solitonic spectra and
partition functions.
On the other hand, the $S_m$ (\ref{S_m}) spectrum is also equal to
that of the loop model in Section 3.1, up to a parameter change,
because they correspond to different choices of cut-off in the same theory.
These facts explain the matching of $Z_{sol}$ for the local
and nonlocal theories, Eqs. (\ref{Z_sol_loc}) and (\ref{Z_sol_t}) 
 (for $M_0=0$).

Two conclusions can be drawn from this analysis:
\begin{itemize}
\item
The on-shell correspondence provides a check for the calculation
of soliton configurations through the $(3+1)$-dimensional extension
of the the loop model.
\item
The IR regularization of the loop model with a fixed mass
parameter $M$ violates scale invariance at the quantum
level, in disagreement with the fermionic dynamics. 
Therefore, another choice of cutoff is needed.
\end{itemize}

Let us consider the  cutoff given by the spatial dimension 
of the system, namely replace:
\be
M\ \longrightarrow\ \frac{1}{\sqrt{R_{1}R_{2}}},
\label{geo_cutoff}
\ee
in the expressions of Section 3.1

Within this choice, the solitonic partition function $Z_{sol}$ 
(\ref{Z_sol_t}) of the loop model becomes:
\begin{multline}
Z_{sol} = \sum_{N_{0},N_{1},N_{2},M_{0}\in \mathbb{Z}} \exp
\biggr\lbrace -\beta \dfrac{1}{\sqrt{R_{1}R_{2}}} \biggr[
\dfrac{2 \pi g_{0}}{k^{2}} \left(
  N_{1}^{2} \dfrac{R_{2}}{R_{1}} +N_{2}^{2} \dfrac{R_{1}}{R_{2}} \right)
 + \dfrac{N_{0}^{2}}{8\pi g_{0}} + \dfrac{M_{0}^{2}g_0}{16\pi^2 q_{0}^{2}}
 \biggr] \biggr\rbrace\, .
\label{Z_osc_t2}
\end{multline}
This expression is manifestly scale invariant and also
invariant under $R_{1} \Leftrightarrow R_{2}$.

Let us remark that the choices of ``geometric cutoff'' in
(\ref{Z_osc_t2}) and ``fixed cutoff'' in (\ref{Z_sol_t}) and (\ref{S_m})
actually amount to two different definitions of the nonlocal
theory at the quantum level.  In the following we adopt the first choice
realizing a scale invariant theory. Further justifications will arise in
 the study of the partition function on the $S^{2} \times
\mathbb{R}$ geometry.

\section{Quantization on the cylinder $S^{2} \times \mathbb{R}$}

In this section, we compute the partition function for the manifold
$S^{2} \times \mathbb{R}$, made by a spatial sphere and
Euclidean time.  As is well-known, this geometry can be mapped to flat space by
the conformal transformation $r=R \exp(u/R)$, where $r$ is the radius
of $\mathbb{R}^3$ and $u$ is Euclidean time on the cylinder.  
It follows that time evolution on the cylinder corresponds to
dilatations in $\mathbb{R}^3$ and the energy spectrum gives access to
conformal dimensions of the fields in the theory \cite{cft} \cite{cc}.
The partition function is schematically:
\be
Z\sim \sum_\D\, \exp\left(-\b\frac{v \D}{R} \right),
\label{Z_conf}
\ee
where $\D$ are the conformal dimensions and $v$ is the Fermi velocity.

The computation of the partition function will follow the same steps
as  in the previous section by
using the four-dimensional formulation.  We consider the manifold
$\mathcal{M}_{4}= S^{3} \times \mathbb{R}$ and embed the
three-dimensional space $\mathcal{M}_{3}= S^{2} \times \mathbb{R}$ by
identifying $S^2$ with the equator of $S^3$.

The four-dimensional  Minkowskian action \eqref{S_loop_4d} on 
$\mathcal{M}_{4} = S^{3} \times \mathbb{R}$ takes the form:
\begin{equation} 
\label{S_4_sph}
S_4[\hat{a}, \zeta] = -\dfrac{1}{4 e^{2}}
\int_{\mathcal{M}_{4}}{dx\,\sqrt{-g} g^{\mu \alpha}g^{\nu \beta}
  \hat{f}_{\mu \nu} \hat{f}_{\alpha \beta}} 
+\dfrac{k}{2\pi}\int_{\mathcal{M}_{3}}{a d\zeta}\, .
\end{equation}

This action is conformal invariant at the classical level:
four-dimensional transformations may induce a nontrivial metric
on ${\cal M}_3$, but this is ineffective on the Chern-Simons action.
Our strategy will be that of assuming conformal invariance
in the quantum theory and then check it in the results
(using the IR cutoff compatible with dilatations).

\subsection{Solitonic modes on $S^2$ embedded in $S^3$} 
\label{sol_sph}

The four-dimensional manifold $S^3\times\mathbb{R}$ is described by
the metric $ds^{2} = dt^{2}-R^2
d\Omega_{3}^{2}$, in terms of $S^3$ polar coordinates,
$d\Omega_{3}^{2}= \sin^2\psi(d\th^2 +\sin^2\th d\vf^2)$, with
$\psi,\th\in [0,\pi]$ and $\vf \in [0,2\pi]$. The $S^2$ sphere
at the equator is identified by $\psi=\pi/2$.

On the geometry of the sphere, there exist
global magnetic fluxes for the $a$ and $\z$ fields. 
These obey, as in (\ref{monod_z}) and (\ref{monod_a}),
\begin{align}
& \int_{S^{2}}{d a}= \dfrac{2 \pi}{q_0} M_{0}\, , \qquad\quad M_0\in\Z,
\label{monod_a_sph}
  \\ \nonumber \\ 
&  \int_{S^{2}}{d\zeta}= \dfrac{2 \pi}{k} N_{0}\, ,\qquad\quad N_0\in\Z.
\label{monod_z_sph}
\end{align}
The electric fluxes for the $a$ field
are instead absent because cycles on $S^2$ are topologically trivial.

Following the same steps as in the previous section, we solve the
equations of motion for the action (\ref{S_4_sph}), 
with source term localized on ${\cal M}_3$. This can be rewritten:
\be
\frac{k}{2\pi}\int_{{\cal M}_4}\d(\mathcal{M}_3)\, ad\z, 
\qquad \delta(\mathcal{M}_{3}) = \dfrac{\delta(\psi-\pi/2)}{R \sin^2(\psi)}\, .
\label{source_sph}
\ee
Note that the form of the delta function is covariant
under translations along the $\psi$ coordinate, i.e. displacements
of $S^2$ from the equator of $S^3$.

The $\z$ magnetic flux (\ref{monod_z_sph}) amounts to a ``charge density''
located at $\psi=\pi/2$ coupled to $\hat{a}_0$ by
the Poisson equation,
\begin{equation} 
\label{poiss_sph}
\nabla_{\mu}\nabla^{\mu} \hat{a}_{0} = 
- \dfrac{e^2 N_{0}}{4\pi R^{2}} 
\dfrac{\delta(\psi-\pi/2)}{R \sin^{2}(\psi)}\, .
\end{equation}
In this equation, it is natural to assume that $\hat{a}_{0}$
depends only on $\psi$, and thus the covariant
Laplacian reduces to an ordinary differential equation.
The solution is easily found to be:
\begin{equation} 
\label{a0_sph}
\hat{a}_{0}(\a) = \dfrac{e^2 N_0}{8\pi R} \left\vert 
\tan(\a) \right\vert\, ,
\qquad\quad \a=\psi-\frac{\pi}{2}\, .
\end{equation}

The other solitonic solution \eqref{monod_a_sph} is a magnetic flux
for the $a$ field that is orthogonal to $S^2$ and can be chosen to be
a constant for all $\psi$ values, i.e. all embeddings $S^2\subset
S^3$:
\begin{equation} 
\label{B_psi}
  B_\psi(\psi)= \hat{f}_{\theta \varphi}(\psi) = \dfrac{M_{0}}{2q_{0}}
  \frac{1}{R^2 \sin^2(\psi)}\, .
\end{equation}

We now compute the energies associated to the two solitonic
solutions (\ref{a0_sph}), (\ref{B_psi}). The Hamiltonian is given by,
\begin{equation} 
\label{H_sph}
H = \dfrac{1}{2e^{2}} \int_{S^{3}}{d^{3}x\, \sqrt{g}
  \left[\hat{f}_{i0}\hat{f}_{j0} g^{ij} + \dfrac{1}{2}
    \hat{f}_{ij}\hat{f}_{lk} g^{il} g^{jk} \right] }\, ,
\end{equation}
where we recognize the electric and magnetic parts.
The electric contribution is obtained by inserting the solution
(\ref{a0_sph}) for $\hat{f}_{\psi 0} = \de_{\psi} \hat{a}_{0}$:
\be
H_{el} = \dfrac{1}{2e^{2}} \int_{S^{3}} d^{3}x\,
\sqrt{g}\, g^{\psi \psi} \left(\de_{\psi} \hat{a}_{0}\right)^{2}
= \dfrac{N_{0}e^{2}}{32 \pi R}
\int_{-\pi/2}^{\pi/2} d \alpha \dfrac{1}{\cos^{2}(\alpha)}\, .
\label{H_el_sph}
\ee  

This integral is divergent at the two poles of $S^3$,
$\a= \pm \pi /2$: an infrared cutoff is again needed.
Let us first introduce a fixed scale, by setting a maximal ``length''
$|R\tan(\a)|<1/(2M)$: we obtain the result,
\be
H_{el} = \dfrac{N_{0}^2}{8g_0 R} \left(\dfrac{1}{MR} \right),
\label{H_el_sph2}
\ee
in terms of the loop model coupling $g_0$ given by (\ref{e_g}).

The magnetic energy is similarly computed from the solution (\ref{B_psi}):
\be
H_{mag} = \dfrac{1}{2e^{2}} \int_{S^{3}} d^{3}x\, \sqrt{g}
  \left(\hat{f}_{\theta \varphi} g^{\theta \varphi}\right)^{2}
  = \dfrac{M_{0}^2 g_0}{8q_{0}^2 R}
\int_{-\pi/2+\d}^{\pi/2-\d}d\alpha \dfrac{1}{cos^{2}(\alpha)}\, .
\ee
This is the same divergent integral of the electric contribution:
once regularized, it yields:
\begin{equation} 
\label{H_mag_sph}
H_{mag} = \dfrac{M_{0}^2g_0}{8 q_{0}^2 R } \left(\dfrac{1}{MR} \right) .
\end{equation}

The values of the classical energies (\ref{H_el_sph}), (\ref{H_mag_sph}) 
determine the solitonic part of the
partition function on the geometry $S^1 \times S^{2}$:
\begin{equation} 
\label{Z_sol_sph}
  Z_{sol} = \sum_{N_{0},M_{0} \in \mathbb{Z}} \exp \left\{
    -\frac{\beta}{R} \left(\dfrac{1}{8MR}\right)
  \left[ \frac{N_{0}^{2}}{g_0} +
    \frac{g_0 M_{0}^{2}}{q_{0}^{2}} \right] \right\}\, .
\end{equation}
We note again that the fixed cutoff $M$ is 
incompatible with scale invariance.
In analogy with the torus case, we replace this scale with
the system dimension, $M=1/(8\l R)$ with $\l$ a numerical constant.
We thus obtain:
\begin{equation} 
\label{Z_sol_sph2}
  Z_{sol} = \sum_{N_{0},M_{0} \in \mathbb{Z}} \exp \left\{
    -\frac{\beta\l}{R} 
  \left[ \frac{N_{0}^{2}}{g_0} + g_0\frac{M_{0}^{2}}{q_0^2} \right] \right\}\, .
\end{equation}
The form of $Z_{sol}$ is now in agreement with conformal invariance,
Eq. (\ref{Z_conf}), and the free parameter $\l$ enters in the definition of
the non-universal Fermi velocity.
The expression (\ref{Z_sol_sph2}) is an important result of our work:
we shall analyze it after completing the derivation
of partition function.

\subsection{Oscillator spectrum}

The oscillator part $Z_{osc}$ is obtained from the
Euclidean $(2+1)$-dimensional action (\ref{S_loop_z2}), by evaluating 
the determinant of the nonlocal kernel.
The action can be rewritten in the form (for $\de_\mu\z^\mu=0)$:
\be
S[\z]=\frac{k^2}{2\pi g_0}\int_{\mathbb{R}_3} 
\z_\mu(x_1) \frac{\d^{\mu\nu}}{(x_1-x_2)^4} \z_\nu (x_2)\, .
\label{S_osc_x}
\ee
Under the conformal map $r=R\exp(u/R)$ from $\mathbb{R}^3$ 
to ${\cal M}_3=\mathbb{R}\times S^2$, with respective
coordinates $x^\mu=(r,\th,\vf)$ and $\wt{x}^\a=(u,\th,\vf)$,
the action is covariant, 
\be
S[\z]=\frac{k^2}{2\pi g_0}\int_{{\cal M}_3} d^3\wt{x}_1d^3\wt{x}_2
\sqrt{g(\wt{x}_1)g(\wt{x}_2)}\ 
\wt{\z}_\a(\wt{x}_1) 
\left(\frac{e^{2(u_1+u_2)/R}}{(x_1-x_2)^4}\right)g^{\a\b}\,
\wt{\z}_\b(\wt{x}_2),
\label{S_osc_sph}
\ee
where the transformations are \cite{cc}, 
$b_\mu dx^\mu=\wt{b}_\a d\wt{x}^\a$, $g^{\a\b}=\d^{\a\b}e^{(u_1+u_2)/R}$,
 and the expression in parenthesis is the correlator of scalar
conformal fields with dimension $\D=2$ on the cylinder. 
Note that the expression (\ref{S_osc_sph}) is
conformal invariant but not reparameterization invariant.

The first step in the calculation of the determinant is that of finding
the eigenvalues: these are obtained by the
spectral decomposition of the $1/x^4$ correlator in the covariant basis
of the cylinder, i.e. Fourier modes $\exp(i\w u)$ and spherical
harmonics $Y_\ell^m(\th,\vf)$. 
Next, the determinant is obtained by zeta-function
regularization of the product of eigenvalues \cite{cc}.
This rather long calculation is done in Appendix B: here we
report the main steps.

The spectral decomposition reads:
\be
\frac{e^{2(u_1+u_2)/R}}{(x_1-x_2)^4}=\frac{8}{R^4}
\sum_{\ell=0}^\infty\sum_{m=-\ell}^{m=\ell}\int_\infty^\infty d\w \, 
e^{i\w(u_1-u_2)}\, Y_\ell^m(\th_1,\vf_1)\, \l_{\w,\ell}\, 
Y^{m*}_\ell(\th_2,\vf_2) ,
\label{4_ker}
\ee
where the eigenvalues are,
\be
\l_{\w,\ell}=\sum_{k=0}^\infty \frac{2k+\ell+2}{(\w R)^2+(2k+\ell+2)^2}
\frac{\G(k+3/2)\G(k+\ell+2)}{\G(k+\ell+3/2)\G(k+1)}\, .
\label{lambda_1}
\ee
The sum in this expression is ultraviolet divergent because $1/x^4$
is not a proper distribution. Rather surprisingly, it can be
evaluated, with result:
\be
\l_{\w,\ell}=\left(\sum_{k=0}^\infty \frac{1}{2}\right) +\frac{\ell+1}{4}
-\frac{\pi}{4}
\left\vert\frac{\G\left((\ell+2 +i \w R)/2\right)}
{\G\left((\ell+1+i \w R)/2\right)} \right\vert^2 \, .
\label{lambda_2}
\ee
The first two terms in this expression, respectively divergent and finite, 
correspond to functions with support for $x_1=x_2$ only, that are subtracted 
for defining the renormalized $1/x^4$ kernel.

The product of eigenvalues can be simplified by using an infinite-product 
representation of the Gamma function; dropping inessential factors, one finds:
\be
\prod_{n\in\Z,\ \ell\ge 0}\l_{n,\ell} \propto 
\prod_{n\in\Z,\ \ell\ge 0}\wh{\l}_{n,\ell} \, , \qquad\quad
\wh{\l}_{n,\ell}= \left(\frac{2\pi n R}{\b}\right)^2 +\L_\ell,
\label{lambda_3}
\ee
where $\l_{n,\ell}=\l_{\w,\ell}$ for discretized momentum
$\w=2\pi n/\beta$ on $S^1$ and $\L_\ell$ refer to angular momentum.
The eigenvalues $\wh{\l}_{n,\ell}$ have now the standard
form of Laplace-type operators on the geometry $S^1\times S^2$.

The regularization of the determinant is obtained by introducing
the zeta-function:
\be 
\z_{S^1\times S^2}(s)=\sum_{\ell=\ell_{min}}^\infty \sum_{n\in\Z} 
\frac{\d(\ell)}{(\wh{\l}_{n,\ell})^s},
\label{Z_func}
\ee
where $\d(\ell)$ is the multiplicity of eigenvalues. The
analytic continuation from large positive
values of $\R (s)$ to $s\sim 0$ leads to the following expression 
of the partition function \cite{cc},
\be
Z_{osc}=\exp\left\{\left.\frac{1}{2}\frac{d}{ds}\z_{S^1\times S^2}(s)
\right\vert_{s=0}\right\}
= e^{-\b{\cal C}/R} \prod_{\ell=\ell_{min}}^\infty\left[ 
1- \exp{\left(-\frac{\b\sqrt{\L_\ell}}{R}\right)} 
\right]^{-\d(\ell)} \, ,
\label{Z_osc_sph}
\ee
where the Casimir energy ${\cal C}/2R$ is 
obtained by evaluating the further zeta-function,
\be
{\cal C}= \z_{S^2}\left(-1/2\right)\, ,
\qquad\quad
\z_{S^2}(s)=\sum_{\ell=\ell_{min}}^\infty \frac{\d(\ell)}{\L^s_\ell}.
\label{casimir_sph}
\ee

The resulting partition function for the loop model takes the
form (\ref{Z_osc_sph}) with parameters 
$({\cal C}, \sqrt{\L_l}, \d(\ell), \ell_{min})$ 
given in the first line of Table \ref{tab1}. The results of
other quadratic theories are also reported in this Table for 
the following discussion.

\section{Conformal invariance and spectrum of the loop model}

In this section, we discuss some interesting informations on 
the spectrum that can be drawn from the expression of $Z=Z_{sol}Z_{osc}$
on $S^1\times S^2$.

\subsection{Particle-vortex duality} 
The solitonic spectrum in $Z_{sol}$ given by (\ref{Z_sol_sph2})
involves ``electric'' and ``magnetic''
quantum numbers $N_0$ and $M_0$, respectively.
In the fermionic case, corresponding to $k=1$ and minimal charge $q_0=1$,
the spectrum is manifestly invariant for $g_{0} \rightarrow 1/g_{0}$. 
This self-duality is expected, because the conformal fields
characterize many observables of the theory and should occur in
self-dual pairs.

On the other hand, the solitonic spectrum on the torus $\mathbb{T}^3$,
given by (\ref{Z_sol_t}) is not self-dual, even for vanishing electric
fluxes $N_1=N_2=0$.  Actually, the $(2+1)$-dimensional duality is not a
symmetry of the partition function, but a Legendre transformation, as
explained in Section 2.4.1. This cannot be verified in our expressions
of $Z$ with vanishing $A$ background: one would need to extend the
derivation for constant $A$, compute the conductivities 
and check that they obey the reciprocity relation (\ref{cond_d}).

\subsection{Conformal invariance}
The conformal invariance of the loop model is rather natural in the
$(3+1)$-dimensional formulation (\ref{S_loop_4d}), as discussed in
Section 4, but is not obvious in the nonlocal form in
$(2+1)$ dimensions (\ref{S_loop_quant}).  The quantization procedure
has actually shown that scale invariance of the solitonic spectrum is
only realized by using a proper IR cutoff.  The oscillator part
$Z_{osc}$ (\ref{Z_osc_sph}) provides further evidences of conformal
invariance at the quantum level:
\begin{itemize}
\item
In a conformal theory, the Casimir energy on $S^2\times\mathbb{R}$ is
related to the trace anomaly, that vanishes in $(2+1)$ dimensions
\cite{cc}. The result ${\cal C}=0$ (cf. Table \ref{tab1})
matches this expectation, as non-zero values would have implied
non-anomalous classical terms in the trace of the stress tensor.
\item
The integer-spaced dimensions of descendent (derivative) fields is
also apparent by the fact that $\sqrt{\L_\ell}\in\Z$ in Table
\ref{tab1}.  For example, the spectrum of non-conformal local
Yang-Mills theory $(2+1)$ dimensions, also reported in the Table,
does not have this property: thus, 
energies do not correspond to scale dimensions, i.e. the 
theory is not covariant under the conformal map to the plane.
\end{itemize}

\subsection{Comparison with other theories}
The loop model corresponds to the large $N$ limit of 
mixed-dimension $QED_{4,3}$: it has a quadratic action but is not a
free theory. The inclusion of solitonic modes makes it an interesting
conformal theory, that is similar to the compactified boson theory
in $(1+1)$ dimensions. The results for the partition function of 
some free conformal theories reported in Table \ref{tab1}
provide other elements for this discussion.

The data indicate that the spectrum of descendent fields is integer
as in $(3+1)$-dimensional theories, while the conformal scalar in
$(2+1)$ dimensions starts from $\D_\vf=1/2$. On the
other hand, the multiplicities $\d (\ell)$ are linear in $\ell$
as in $(2+1)$ dimensions, instead of being quadratic,
a characteristic feature of angular momentum on $S^3$.

Going back to the $(3+1)$-dimensional action (\ref{S_loop_4d}) and integrating
over the $\z$ field, one find that the loop model can be seen as a
constrained Yang-Mills theory, enjoying a subspace of its Hilbert
space. The comparison between the first and last lines of Table
\ref{tab1} shows this fact.  
In conclusion, the loop model is a conformal theory with
mixed-dimension properties, whose features would need a deeper
analysis using representation theory of the conformal group.

\begin{table}
\centering
\be 
\begin{array}{|l|c|c|c|c|c|}
\hline  
{\rm Theory} & {\rm dimension}& {\cal C} & \sqrt{\L_\ell} & \d(\ell) 
& \ell_{min} \\
\hline
{\rm loop\ model}   &  (2+1)    & 0    & \ell & 2\ell &  1 \\
{\rm conformal\ scalar} & (2+1) & 0    & \ell+\frac{1}{2} & 2\ell+1 & 0 \\
{\rm vector}\quad   & (2+1) & \neq 0& \sqrt{\ell(\ell+1)} & 2\ell+1 & 1\\
{\rm conformal\ scalar} & (3+1) &  \frac{1}{120} & \ell & \ell^2 & 1 \\
{\rm vector}\quad   & (3+1) & \frac{3}{20} & \ell & 2(\ell^2-1)& 2\\
\hline
\end{array}
\nonumber
\ee
\caption{Parameters entering in the partition function (\ref{Z_osc_sph}) of
some quadratic theories \cite{cc}: Casimir energy $\cal C$; energy level
$\sqrt{\L}_\ell$; eigenvalue degeneracy $\d(\ell)$; minimal 
value $\ell_{min}$}.
\label{tab1}
\end{table}

\subsection{Anyon excitations}

Let us analyze the results of Section 4 for $k>1$, that are
relevant for the dynamics at the surface of interacting topological
insulators (cf. Section 2.3).
In this case, the partition function (\ref{Z_sol_sph}) should describe 
excitations with fractional charge and statistics in $(2+1)$ dimensions.
The subject is well understood for non-relativistic dynamics, as
e.g. in the fractional quantum Hall effect. 
The loop model provides a description in
the relativistic scale-invariant domain.

The form of the surface action (\ref{S_surf_loop}) in Section 2.3.4,
\be
S_{surf}[a,\z,A]=\frac{i}{2\pi}\int \left( k\z da +\z dA \right)+ S_{loop}[a],
\label{S_surf_simpl}
\ee
tells us that:
\begin{itemize}
\item
The $\z$ field is dual to the background $A$ with minimal charge 
$e_0=1/k$, Eq. (\ref{min_e}); thus,  magnetic excitations of $\z$
possess minimal charge $\wt{e}_0=1/k$ in agreement with the
quantization condition (\ref{monod_z}).
\item
The $a$ field is dual to $\z$, i.e. it is electric, and possesses
minimal charge $q_0=1$, as confirmed by the constraint $A\sim k\, a$
implemented by $\z$.  Therefore, its monopoles have minimal charge
one  for any $ k$ value (cf. Eq.(\ref{monod_a}) for  $q_0=1$).
\item
The map between the actions (\ref{S_surf_loop}) and (\ref{S_surf_z}), 
i.e. by integrating the $a$ field, is a generalization of the particle-vortex
duality transformation for theories with fractional charges (cf.
Section 2.4.1).  In this transformation, the loop-model coupling is
mapped into:
\be 
\wt{g}_0\equiv g=\frac{k^2}{g_0}.
\label{k_duality}
\ee
\end{itemize}

These results lead us to consider the solitonic spectrum (\ref{Z_sol_sph2})
at the electric-magnetic self-dual point $g_0=k$:
\be
E_{sol}=\frac{v}{R}\D_{N_0,M_0}\, ,\qquad
\D_{N_0,M_0}=\frac{1}{2}\left[\frac{N_0^2}{k} +k M_0^2\right],
\qquad\quad (g_0=k).
\label{e_sol_spec}
\ee
Upon writing $N_0=kn +m$, with $m=0,1,\dots, k-1$ and $n\in \Z$,
this spectrum contains states with fractional dimensions 
$\D=m^2/(2k)+\Z$. Thus, there are $k$ independent anyonic sectors in
agreement with the value $k$ of
the topological order on the $S^2\times S^1$ geometry (this can be
computed from the bulk BF theory, as explained e.g. in 
Section 3.3.1 of Ref.\cite{CRS}).

Furthermore, the behaviour of conformal correlators 
 on the surface of topological insulators  should match
the known Aharonov-Bohm phases between excitations predicted 
by the BF theory (\ref{S_BF}), 
\be
\th=\frac{2\pi n_1 n_2}{k}, \qquad\quad n_1,n_2\in \Z\, .
\label{any_monod}
\ee
Let us explain this point in some detail.

As nicely discussed in Ref.\cite{marino}, order-disorder fields in
$(2+1)$ dimensions require: i) gauge fields and ii) a
symplectic structure. Given the equal-time commutation relations,
\be
\left[a_i(x,t),\pi^j (y,t)\right]=i\d_i^j\d^{(2)}(x-y), \qquad\quad
i,j=1,2,
\label{simpl}
\ee
between the gauge field $a$ and its conjugate momentum $\pi$,
the order and disorder operators take the form, respectively,
\ba
\s(x,t)&=&\exp\left(-i\a \int_{-\infty}^x d\xi^i a_i(\xi,t) \right),
\nl 
\mu(x,t)&=& \exp\left(i\b \int_{-\infty}^x 
d\xi^i \eps_{ij} \pi^j (\xi,t)\right),
\label{o_d}
\ea
where the line integrals go to $-\infty$ along a given common direction,
e.g. the negative real axis. Upon using the identity 
$\eps^{ij}\de_i\de_j {\rm Arg}(x-y)=\pi \d^{(2)}(x-y)$, 
one finds the (equal-time) monodromy:
\be
\mu\left(e^{i2\pi}z,t\right)\s(0,t)=e^{i2\a\b}\, \mu(z,t)\s(0,t),
 \qquad\qquad z=x_1+ix_2\, .
\label{o_d_monod}
\ee

This topological information is contained in the part $\int k\,\z da$ of the
action (\ref{S_surf_simpl}), where the canonical momentum is
$\pi^i=k/(2\pi)\eps^{ij}\z_j$, as explained in Section 2.3.3.
Therefore, exponentials
of line integrals (\ref{o_d}) of the $a$ and $\z$ fields realize
the expected monodromies (\ref{any_monod}) at the surface of the topological
insulators, by suitably choosing the $\a,\b$ parameters.

The dynamics introduced by $S_{loop}$ in
(\ref{S_surf_simpl}) yields two-point functions of conformal fields,
$\langle\f(x)\f(0) \rangle=(x^2)^{\D} $. 
Evaluated at equal time, $x_\mu=(0,x_1,x_2)$, the power-law behavior
$|z|^{2\D}$  should match the monodromy phase
(\ref{any_monod}) for reconstructing the analytic dependence $z^{2\D}$ of 
conformal invariance in the two-dimensional plane.
The values of $\D$ in the spectrum (\ref{e_sol_spec}) 
do verify this requirement.

In conclusion, the loop model action (\ref{S_surf_loop}) describes the
surface excitations of fractional topological insulators for the
self-dual value of the coupling constant $g=k$.  The identification of
the conformal spectrum (\ref{e_sol_spec}) also requires a choice of
Fermi velocity $v$.

We remark that the $(1+1)$-dimensional chiral boson theory describing
topological insulators in one lower dimension also involves some
tuning of parameters \cite{cdtz} \cite{cft}. 
Note also that the $(1+1)$-dimensional conformal spectrum,
\be
\D_{n,\ov{n}, m}=\frac{1}{4k}\left[\left(k(n+\ov{n})+2m\right)^2
+\left(k(n-\ov{n})\right)^2 \right],
\label{spec_2d}
\ee
cannot be written in the form (\ref{e_sol_spec}) for odd $k$. Actually,
the $(1+1)$-dimensional theory involves pairing of chiral-antichiral
excitations for respecting time-reversal symmetry, while each
$(2+1)$-dimensional excitation is symmetric.


\section{Conclusions}

In this paper, we have shown that the loop model is a conformal 
theory in $(2+1)$ dimensions that bears some
similarities with the compactified boson in $(1+1)$ dimensions \cite{cft}.
Its coupling constant spans a critical line along which the spectrum
displays fermionic and anyonic excitations, thus providing a
viable approach towards bosonization of free and interacting fermions.
The formulation as a local theory in $(3+1)$ dimensions allows
for other interesting developments.

Let us mention possible extensions of our work:
\begin{itemize}
\item
The generalization of the analysis in presence of the Chern-Simons
interaction (coupling $f\neq 0$ in (\ref{S_loop_x})) will provide a
dyonic spectrum that breaks parity and time-reversal symmetry and is
covariant under more general duality transformations \cite{gaiotto}.
\item
The analysis of order-disorder fields can be extended beyond the simple
observations of Section 5.4. In this respect, we note that in
the $(2+1)$-dimensional formulation (\ref{S_loop_quant}), one gauge field
is non-dynamic or can be integrated out, Eq. (\ref{S_surf_z}).
Thus, either the order or the disorder fields should become
collective excitations.
\item
The loop model can be made interacting by including $1/N_F$
corrections stemming from the relation with $QED_{4,3}$. In this
respect, it provides a viable platform for quantitative discussions of
the dualities  and other interesting aspects of
$(2+1)$-dimensional physics.
\item
Finally, the $(3+1)$ local formulation of the theory can be useful for
studying non-Abelian generalizations.
\end{itemize}

{\bf Acknowledgments}

We would like to thank Z. Komargodski, N. Magnoli, E. Marino,
C. Morais-Smith, G. Palumbo, D. Seminara and P. Wiegmann for very
useful scientific exchanges. We also acknowledge the hospitality and
support by the Simons Center for Geometry and Physics, Stony Brook,
Nordita, Stockholm, and the G. Galilei Institute for Theoretical
Physics, Arcetri. This work is supported in part by the Italian
Ministery of Education, University and Research under the grant PRIN
2017 ``Low-dimensional quantum systems: theory, experiments and
simulations.''

\appendix

\section{Peierls argument}

We evaluate the Euclidean action of the loop model \eqref{S_peierls}
on the configuration of a monopole with minimal magnetic charge $2\pi/q_0$:
\begin{equation} \label{Monopolo}
F_{\mu \nu} = \dfrac{1}{2 q_0}\varepsilon_{\mu \nu \rho}
\dfrac{x_{\rho}}{|x|^{3}}\, .
\end{equation}
The integral of the nonlocal term in
\eqref{S_peierls} reads:
\ba
S & =& \dfrac{g}{32\pi^{3} q_0^2}
\int{\dfrac{d^{3}x_{1}}{|x_{1}|^{3}}\,
  \dfrac{d^{3}x_{2}}{|x_{2}|^{3}}\, \dfrac{(x_{1} \cdot
    x_{2})}{|x_{1}-x_{2}|^{2}} } 
\nl
&=& \dfrac{g}{4 \pi q_0^2} \int_0^\infty d\a \int_{0}^{\infty}{dr_{1}}
\int_{0}^{\infty}{dr_{2}} \int_{-1}^{1}{dy\, } y\,
e^{-\a({r_{1}^{2}+r_{2}^{2}-2r_{1}r_{2}y})} \, ,
\ea
where we have used polar coordinates, exponentiated the denominator
and introduced the variable $y=\cos(\th_1-\th_2)$. 
Upon rescaling the radii, $s_i=r_i \sqrt{\a}$, $i=1,2$,
the integral factorizes into a logarithmic divergent 
part and a finite part,
namely the integrals over $\a$ and over the others variables.

We observe that being $\a$ conjugated to $r^2$, 
we can regularize the divergent contribution as follows:
\begin{equation} 
\int_{0}^{+\infty} \dfrac{d\alpha}{\alpha} \quad\rightarrow\quad
\int_{1/L^{2}}^{1/a^{2}} \dfrac{d\alpha}{\alpha}= 2 \ln \biggr(
\dfrac{L}{a} \biggr) \, ,
\end{equation}
where $\a$ and $L$ are the lattice constant and the system size
respectively. On the other hand the finite part can be evaluated in
polar coordinates $s_{1} = s \cos(\eta)$, $s_2=s \sin(\eta)$, 
leading to the result \eqref{deltaF_mono}.

\section{Loop-model determinant on  $S^{2}\times \mathbb{R}$ }

In this appendix we give some details concerning the calculation of
the oscillator spectrum and determinant of the loop model reported in
Section 4.2. The first step is the spectral decomposition of the $1/x^4$
kernel in the action (\ref{S_osc_x}).

\subsection{Kernel decomposition}
As a warming up, we determine the spectral form of the
propagator of scalar fields,
\be
\la \f(x_1)\f(x_2)\ra _{\mathbb{R}^3} = \dfrac{1}{\vert{x_1-x_2}\vert}.
\label{}
\ee
The conformal map from flat space $x^\mu=(r,\th,\vf)$ 
to the cylinder $\wt{x}^\a=(u,\th,\vf)$ is obtained by transforming the
fields, $\widetilde{\f}=e^{ u/2R}\f$, leading to:
\be
\la \widetilde{\f}(\wt{x}_1)\widetilde{\f}(\wt{x}_2)
\ra _{\mathbb{R}\times S^2}=\frac{e^{(u_1+u_2)/2R}}{\vert x_1-x_2 \vert}.
\ee 

This expression can be expanded in terms of Legendre polynomials $P_\ell$
and spherical harmonics $Y_\ell^m$, by using \cite{harmonic}:
\ba
\frac{1}{\vert x_1-x_2 \vert} &=&
\sum_{\ell=0}^{\infty}\dfrac{r_1^\ell}{r_2^{\ell+1}}
P_{\ell}\left(\widehat{x}_1\cdot\widehat{x}_2\right), \qquad\quad
x_i=r_i\wh{x}_i,\ \ i=1,2,\ \
r_1<r_2,
\nl
P_{\ell}\left(\widehat{x}_1\cdot\widehat{x}_2\right)&=&\frac{4\pi}{2\ell+1}
\sum_{m=-\ell}^\ell Y_{\ell}^{m\,*}\left(\th_1,\vf_1\right)
Y_{\ell}^{m}\left(\th_2,\vf_2\right)\, .
\ea
Introducing the Fourier modes $e^{i\w u}$, we obtain the spectral decomposition:
\be
\begin{aligned}
\la \widetilde{\f}(x_1)\widetilde{\f}(x_2)\ra _{\mathbb{R}\times S^2}
&=4\pi\int_{-\infty}^{\infty} d\w\,\sum_{\ell=0}^{\infty}
\sum_{m=-\ell}^{\ell}e^{i \w ( u_1-u_2) }\,
Y_{\ell}^{m\,*}\left(\th_1,\vf_1\right)\,\l_{\w,\ell}\,
Y_{\ell}^{m}\left(\th_2,\vf_2\right)\, ,
\\
\l_{\w,\ell} &=
\dfrac{1}{\left(R \w\right)^2 +\left(\ell+1/2\right)^2}.
\end{aligned}
\ee
This spectrum  confirms that the propagator is 
the inverse of the conformal Laplacian  in $(2+1)$ dimensions,
as reported in Table 1 for the conformal scalar theory.

Let us now apply the same procedure to the $1/x^4$ kernel. 
We use the identity,
\be
\dfrac{1}{\vert x \vert^4}=\dfrac{1}{2\vert x \vert}
\int_{0}^{\infty}dp\,p^2\, e^{-p\vert x \vert},
\ee
and the formula:
\be
\dfrac{e^{-p\vert x_1 - x_2\vert}}{\vert x_1 - x_2\vert}=
\sum_{\ell=0}^{\infty}\dfrac{\left(2\ell+1\right)}{\sqrt{r_1 r_2}} 
I_{\ell+\frac{1}{2}}\left(pr_1\right)K_{\ell+\frac{1}{2}}\left(pr_2\right)
P_{\ell}\left(\widehat{x}_1\cdot\widehat{x}_2\right),\qquad\quad r_1<r_2,
\ee
where $I_{m}$ and $K_{m}$ are modified Bessel functions of 
the first and second kind, respectively.
The integration over $p$ of the Bessel functions leads to
the Hypergeometric function ${_2 F_1}$; the 
kernel with appropriated Weyl factors is then written:
\be
\dfrac{e^{2\left(u_1+u_2\right)/R}}{\vert x_1-x_2\vert^4}
=\dfrac{\sqrt{\pi}}{R^4}\sum_{\ell=0}^{\infty}
e^{-\vert u\vert(\ell+2)/R}
\dfrac{\G\left(\ell+2\right)}{\G\left(\ell+\frac{1}{2}\right)}\, 
{_2 F_1}\left(\dfrac{3}{2},\ell+2,\ell+\frac{3}{2} ; e^{-2\vert u\vert}\right)
P_\ell\left(\widehat{x}_2\cdot\widehat{x}_2\right),
\label{ker_four}
\ee
where $u=u_1-u_2$. 
Finally, the series expansion of the Hypergeometric function allows
one to compute the Fourier modes, leading to the spectral decomposition
(\ref{4_ker}) with eigenvalues \eqref{lambda_1}:
\be
\l_{\w,\ell}=\sum_{k=0}^\infty \frac{2k+\ell+2}{(\w R)^2+(2k+\ell+2)^2}
\frac{\G(k+3/2)\G(k+\ell+2)}{\G(k+\ell+3/2)\G(k+1)}\, .
\label{lambda_4}
\ee

\subsection{Field decomposition}

The spin-one field on the cylinder $\wt{\z}$ is expanded in the basis
of vector spherical harmonics ${\rm Y}^{JL S M}_\mu$, with $S=1$, that can be
written in terms of scalar harmonics $Y^m_L$ and constant vectors
$\chi_\mu^m$ by using the addition of angular momenta \cite{harmonic}:
\be
\begin{aligned}
\widetilde{\z}_\m \left(\wt{x}\right)&=\int \dfrac{d\w}{2\pi}\, e^{i\w u}
 \sum_{J=1}^{\infty}\sum_{L=J-1}^{J+1}\sum_{M=-J}^{J}
\wt{\z}_{J,L,M}\left(\w\right){\rm Y}_{\mu}^{JL1M}\left(\th,\vf\right)\, ,
\\
{\rm Y}_{\mu}^{JL1M}\left(\th,\vf\right)&=
\sum_{m=-L}^{L}\sum_{m'=-1}^{1}C_{L,1}\left(J,M,m,m'\right)
Y_{L}^{m}\left(\th,\vf\right)\chi_{\mu}^{m'},
\end{aligned}
\label{}
\ee 
where $C_{L,1}\left(J,M,m,m'\right)$ are the Clebsh-Gordan coefficients
with $M=m+m'$.

Upon substituting the previous expansions in the
Euclidean action (\ref{S_osc_x}) and making use of orthonormality,
we obtain:
\be
S[\z]\propto\int \dfrac{d\w}{2\pi}\,\sum_{L=0}^{\infty}
\sum_{J=L-1}^{L+1}\sum_{M=-J}^{J}\left\vert \wt{\z}_{J,L,M}
\left(\w\right)\right\vert^2\, \l_{\w,L}\, ,
\label{f_exp}
\ee
where $\wt{\z}_{-1,L,M} \left(\w\right)=0$.
The eigenvalues $\l_{\w,L}$ (\ref{lambda_4}) of the scalar kernel
(\ref{ker_four}) only  depends on the orbital momentum and reduce the 
summations in (\ref{f_exp}) to a single one over $L=0,1,\dots$, with
multiplicities $\d(L)$. The gauge condition $\de^\m\z_\m=0$ imposes
$\wt{\z}_{L,L,M}(\w)=0$, and one finds,
\be
\d(L)=2\left(2L+1\right)\, .
\ee

\subsection{Resummation and regularization}

The sum over $k$ in the eigenvalues $\l_{\w,\ell}$ \eqref{lambda_4}
is regularized by subtracting the asymptotic $k\to\infty$ limit of the
summand, equal to $1/2$:
\be
I\left(\ell\right)=\sum_{k=0}^{\infty}\left[
\dfrac{\left(\ell+2+2k\right)}{a^2+\left(\ell+2+2k\right)^2}
\dfrac{\G\left(\ell+2+k\right)\G\left(k+3/2\right)}{\G\left(\ell+3/2+k\right)
  \G\left(k+1\right)}-\dfrac{1}{2}\right],
\label{reg_sum}
\ee
where $a=\w R$. The series (\ref{reg_sum}) can be summed by using
the Sommerfeld-Watson method and the result is expressed in terms of
two finite products for even and odd $\ell$ values, respectively:
\begin{equation}
I\left(\ell\right)=\dfrac{\ell+1}{4}-\begin{cases}
	& \ \qquad \dfrac{\pi a}{8}
\displaystyle{\prod_{i=0}^{\left(\ell-1\right)/2}}
\dfrac{\left(2i+1\right)^2+a^2}{\left(2i\right)^2+a^2}
\tanh\left(\dfrac{a\pi}{2}\right)\, , \quad \ell=1,3,\dots\, , \\ \\
	& \displaystyle{\frac{\pi(1+a^2)}{8a} \prod_{i=0}^{\ell/2}}
\dfrac{\left(2i\right)^2+a^2}{\left(2i-1\right)^2+a^2}
\coth\left(\dfrac{a\pi}{2}\right)\, , \quad \ell=0,2,\dots\, .
	\end{cases}
\end{equation}
 Both products are rewritten as a ratio of complex gamma functions squared,
leading to the regularized eigenvalues,
\be
\l_{\w,\ell}^{reg}=I\left(\ell\right)-\dfrac{\ell+1}{4}=-\dfrac{\pi}{4}
\left\vert\frac{\G\left((\ell+2 +i \w R)/2\right)}
{\G\left((\ell+1+i \w R)/2\right)} \right\vert^2 \, ,
\ee
reported in \eqref{lambda_2}. For compact time $\b$, the
Fourier modes are discretized, $\w R=n/\t$, with $\t=\b/(2\pi R)$, $n\in\Z$.

Next, the infinite-product representation of the gamma function
\cite{bateman},
\be
\frac{\G\left(a+ib\right)}{\G\left(a\right)}=e^{-i\g b}
 \left( 1+i \frac{b}{a} \right)^{-1}
\prod_{k=1}^{\infty} e^{ib/k}
\left(1+i\frac{b}{a+k}\right)^{-1}, \qquad\quad a,b\in \mathbb{R},
\ee
is used to rewrite the product of eigenvalues occurring in the
determinant. Dropping inessential $\t$-independent factors,
we obtain the expression:
\ba
\sum_{n,\ell}\d(\ell)\log\left(\l_{n,\ell}^{reg}\right)
&=&\sum_{n\in\Z}\sum_{\ell,k=0}^{\infty}\d(\ell)
\log\left[
\dfrac{n^2 +\t^2(\ell+2k+1)^2}{n^2+\t^2(\ell+2k+2)^2}
\right]
\nl
&=&\sum_{n\in\Z}\sum_{L=0}^\infty 2L \log\left(n^2+\t^2 L^2 \right)\, .
\ea
The sums in this expression simplify because the indices $\ell$ and $k$ 
come in the combination $L=\ell+2k$. The resulting
sum over $n,L$, with multiplicity  $\d(L)=2L$, 
 can now be analytically continued by 
using the zeta-function method, as described in the main text.


\end{document}